\documentclass{article}

\usepackage{microtype}
\usepackage{listings}
\usepackage{graphicx}
\usepackage{subfigure}
\usepackage{booktabs} % for professional tables
\usepackage{mathtools}
\usepackage{etoolbox}
\usepackage{xspace} 
\usepackage{caption}
\usepackage[hyphens]{url}
\usepackage{hyperref}
\usepackage{enumitem}
\usepackage{xcolor}
\usepackage{color}
\usepackage[T1]{fontenc}
\usepackage[scaled=0.9]{sourcecodepro}
\usepackage{makecell}
\usepackage{booktabs}
\usepackage{diagbox}

\usepackage[accepted]{mlsys2024}

\newcommand{\sys}{FlashInfer\xspace}
\newcommand{\papertitle}{\sys: Efficient and Customizable Attention Engine for LLM Inference Serving}
\DeclarePairedDelimiter\ceil{\lceil}{\rceil}

\mlsystitlerunning{\papertitle}

{\ignorespaces}
{\ignorespaces}
{\ignorespaces}
{\ignorespaces}
{\ignorespaces}
{\ignorespaces}
{\ignorespaces}

\definecolor{deepblue}{rgb}{0,0,0.5}
\definecolor{light-gray}{gray}{0.95}
\definecolor{mygreen}{rgb}{0,0.6,0}
\definecolor{unpred-color}{rgb}{0, 0.45, 0.69}
\definecolor{pred-color}{rgb}{0.87, 0.56, 0.19}
\definecolor{incorrect-color}{rgb}{0, 0.61, 0.45}
\definecolor{commentgreen}{RGB}{2,112,10}
\definecolor{eminence}{RGB}{108,48,130}
\definecolor{weborange}{RGB}{255,165,0}
\definecolor{frenchplum}{RGB}{129,20,83}

\lstdefinestyle{zihao-style}{
    keywordstyle=\color{darkgray}\bfseries,
    morekeywords={with},
    basicstyle=\fontfamily{SourceCodePro-TLF}\selectfont\scriptsize\linespread{0.8},
    commentstyle=\color{commentgreen},
    backgroundcolor=\color{light-gray},
    language=Python,
    frame=tlbr,
    framesep=0.1cm,
    framerule=0pt,
    aboveskip=0.3\medskipamount,
    belowskip=0.3\medskipamount,
    keywordstyle=\bfseries\color{deepblue},
    breakatwhitespace=false, % sets if automatic breaks should only happen at whitespace
    breaklines=true, % sets automatic line breaking
    captionpos=b, % sets the caption-position to bottom
    numbers=none, % where to put the line-numbers; possible values are (none, left, right)
    tabsize=2, % sets default tabsize to 2 spaces
    mathescape=true, % allow $x$ in listings
    upquote=true,
    xrightmargin=0.1cm,
    xleftmargin=0.1cm,
}

\begin{document}

\twocolumn[
\mlsystitle{\papertitle}

% unless you've provided the [accepted] option to the mlsys2024


\mlsyssetsymbol{equal}{*}
\mlsyssetsymbol{nv-intern}{*}

\begin{mlsysauthorlist}
\mlsysauthor{Zihao Ye}{nv-intern,uw,nv}
\mlsysauthor{Lequn Chen}{ppl}
\mlsysauthor{Ruihang Lai}{cmu}
\mlsysauthor{Wuwei Lin}{nv}
\mlsysauthor{Yineng Zhang}{independent}
\mlsysauthor{Stephanie Wang}{uw}
\mlsysauthor{Tianqi Chen}{nv,cmu}
\mlsysauthor{Baris Kasikci}{uw}
\mlsysauthor{Vinod Grover}{nv}
\mlsysauthor{Arvind Krishnamurthy}{uw}
\mlsysauthor{Luis Ceze}{uw,nv}
\end{mlsysauthorlist}

\mlsysaffiliation{uw}{Paul G. Allen School of Computer Science \& Engineering, University of Washington}
\mlsysaffiliation{cmu}{Carnegie Mellon University}
\mlsysaffiliation{nv}{NVIDIA}
\mlsysaffiliation{ppl}{Perplexity AI}
\mlsysaffiliation{independent}{Independent Researcher}

\mlsyscorrespondingauthor{Zihao Ye}{zhye@cs.washington.edu}

\mlsyskeywords{Machine Learning, MLSys}

\vskip 0.3in

\begin{abstract}

Transformers, driven by attention mechanisms, form the foundation of large language models (LLMs). As these models scale up, efficient GPU attention kernels become essential for high-throughput and low-latency inference. Diverse LLM applications demand flexible and high-performance attention solutions.
We present \sys: a customizable and efficient attention engine for LLM serving. \sys tackles KV-cache storage heterogeneity using block-sparse format and composable formats to optimize memory access and reduce redundancy. It also offers a customizable attention template, enabling adaptation to various settings through Just-In-Time (JIT) compilation. Additionally, \sys's load-balanced scheduling algorithm adjusts to dynamism of user requests while maintaining compatibility with CUDAGraph which requires static configuration.
\sys have been integrated into leading LLM serving frameworks like SGLang, vLLM and MLC-Engine. Comprehensive kernel-level and end-to-end evaluations demonstrate \sys's ability to significantly boost kernel performance across diverse inference scenarios: compared to state-of-the-art LLM serving solutions, 
\sys achieve 29-69\% inter-token-latency reduction compared to compiler backends for LLM serving benchmark, 28-30\% latency reduction for long-context inference, and 13-17\% speedup for LLM serving with parallel generation. \newline

\textbf{Project page:} \url{http://flashinfer.ai}

\end{abstract}

]
\printAffiliationsAndNotice{\textsuperscript{*}Part of the work was done while Zihao Ye was interning at NVIDIA.}

\section{Introduction}
\label{sec:intro}

The Transformer architecture has become the primary backbone for large language models (LLMs), prominently featuring attention mechanism~\cite{vaswani2017transformer} as its most salient component. As LLMs rapidly evolve and find applications in diverse fields, the demand for efficient GPU attention kernels grows, with the goal of enabling scalable and responsive model inference. At the heart of LLM inference lies the attention computation, which plays a crucial role in processing historical context and generating outputs based on query vectors. In LLM serving, the attention mechanism reads from the KV cache, which stores historical context, and computes outputs based on the current query. The efficiency of this attention operator is paramount to the overall performance of an LLM inference systems. However, creating high-performance attention kernels tailored for LLM serving introduces challenges not typically encountered in traditional training environments.

\begin{figure*}[t!]
    \centering
    \includegraphics[width=0.85\textwidth]{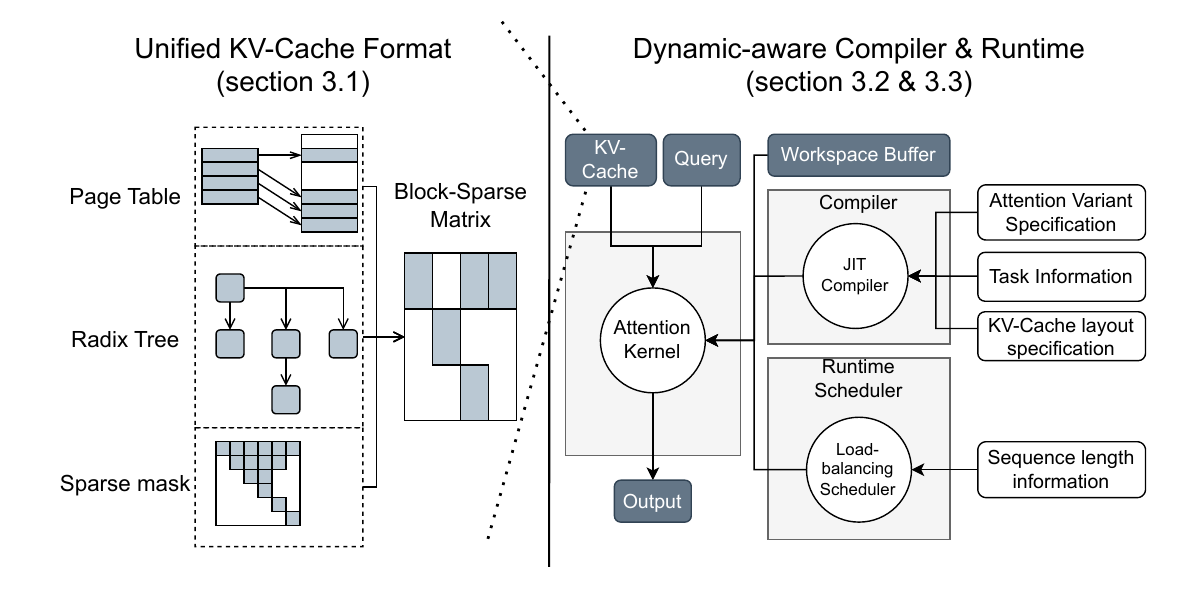}
    \caption{Overview of the \sys system design: Attention variant specifications, task information and KV-cache layout specifics are provided at compile time for JIT compilation, while sequence length information is input at runtime for dynamic scheduling.}
    \label{fig:overview}
\end{figure*}

Two major challenges arise when building efficient attention support for LLM systems:

\textbf{LLM applications exhibit diverse workload patterns and input dynamics.} LLM serving involves various attention computation patterns, from prefill computation for context processing to batched decoding during serving~\cite{yu2022orca}. As multiple requests are processed, opportunities for \emph{prefix-reuse} emerge, and the introduction of tree decoding in speculative scenarios creates additional attention patterns~\cite{cai2024medusa,miao2023specinfer,chen2024sequoia}. Moreover, query lengths and KV caches vary within batches and over time, naive implementation might suffer load-imbalance issue,
optimal scheduling requiring kernel to adapt dynamically for optimal performance.

\textbf{Modern hardware implementations necessitate the customization of attention operators.} On the memory side, efficient storage formats, such as paged attention~\cite{kwon2023vllm} and radix trees~\cite{zheng2023sglang}, are critical for managing the growing KV cache sizes and diverse storage patterns. On the compute side, crafting hardware-specific pipelines and templates is indispensable to fully exploit the performance potential of each GPU architecture~\cite{dao2023flashattention2,jay2024flashattention3}. Furthermore, the design must accommodate the increasing variety of attention mechanisms in modern LLMs, such as grouped attention heads~\cite{ainslie2023gqa, shazeer2019mqa}, specialized masks~\cite{beltagy020longformer}, and customized attention score computations~\cite{morgane2024gemma2,grok,jason2024flashsigmoid}, necessitating flexible and scalable implementation strategies.

The combined complexity of workload diversity and hardware heterogeneity complicates the development of a comprehensive attention solution. Currently, each system implements a specialized attention solution based on a subset of these characteristics, leading to high maintenance overhead and potential inefficiencies. To address these challenges, we introduce \sys, a code-generation based attention engine designed to accelerate attention computation in LLMs. Our approach incorporates several key designs:

\textbf{\sys utilizes a block-sparse format to tackle KV-Cache storage heterogeneity.} This format serves as a unified data structure for various KV-Cache configurations, with adjustable block sizes allowing fine-grained sparsity, such as vector-level sparsity~\cite{chen2021tensorcoressparsity, li2022magicube}. This approach unifies diverse KV-Cache patterns and enhances memory access efficiency.

\textbf{A customizable attention template supports different attention variants in \sys.} \sys provides a customizable programming interface for users to implement their attention variants. \sys uses Just-In-Time (JIT) compilation to translate these variants into highly optimized block-sparse implementations, ensuring rapid adaptation to varying attention configurations.

\textbf{\sys employs a dynamic load-balanced scheduling framework to handle input dynamism effectively.} It separates compile-time tile size selection from runtime scheduling, offering lightweight APIs that adaptively manage scheduling with changing KV-Cache lengths during inference, while maintaining compatibility with CUDAGraph's requirement for constant configurations~\cite{alan2019cudagraph, vinh2021pytorchcudagraph}.

Figure \ref{fig:overview} depicts our system design. We evaluated \sys' performance across standard LLM serving environments and innovative scenarios, including prefix sharing and speculative decoding. \sys have been integrated with mainstream LLM serving engines, including vLLM~\cite{kwon2023vllm}, MLC-Engine~\cite{mlcllm, lai2023relax}, and SGLang~\cite{zheng2023sglang}, we assessed its impact on end-to-end latency and throughput improvements, showing significant enhancements on standard LLM serving benchmarks and novel applications such as long-context inference and parallel generation.

Our contributions include:
\begin{itemize}[itemsep=0pt,topsep=0pt,parsep=0pt]
    \item Introduction of flexible block-sparse and composable formats addressing KV-Cache storage heterogeneity for efficient memory management and access.
    \item Development of a customizable attention template accommodating diverse attention variants, ensuring high-performance execution via JIT compilation.
    \item Design of a dynamic scheduling framework managing input dynamism while remaining compatible with CUDAGraph, maximizing hardware utilization.
    \item Comprehensive evaluation demonstrating substantial improvements in kernel and end-to-end performance.
\end{itemize}
\vspace{-1em}
\section{Background}
\label{sec:background}

\subsection{FlashAttention}
\label{sec:flashattention}

FlashAttention~\cite{dao2022flashattention} is an efficient algorithm for computing exact attention with reduced memory usage. During the forward pass, it employs the online-softmax trick~\cite{milakov2018onlinesoftmax}, updating attention outputs on-the-fly using a constant amount of on-chip memory, thus avoiding materializing the attention matrix in GPU global memory. FlashAttention2\&3~\cite{dao2023flashattention2,jay2024flashattention3} improve performance by optimizing loop ordering and pipeline design for Ampere and Hopper GPUs. \sys builds upon these advancements.

The operational intensity of FlashAttention is given by $O\left(\frac{1}{1/l_{qo} + 1/l_{kv}}\right)$, where $l_{qo}$ and $l_{kv}$ are the query and key-value cache lengths, respectively. In LLM serving, the query length is either equal to (prefill) or smaller than (decode/incremental prefill) the key-value cache length, simplifying the operational intensity to $O(l_{qo})$. Techniques like batching~\cite{yu2022orca} do not alter this operational intensity.
Multi-Query Attention (MQA) ~\cite{shazeer2019mqa} and Grouped Query Attention (GQA) ~\cite{ainslie2023gqa} optimize the KV-Cache size by grouping queries and sharing the same KV-Cache entries. The ratio of the number of queries to the number of KV-Cache entries is denoted as the group size $g = \frac{H_{qo}}{H_{kv}}$, enhancing operational intensity to $O(g \cdot l_{qo})$.

\subsection{Attention Composition}
\label{sec:attention-composition}

Block-Parallel Transformer (BPT)~\cite{liu2023blockparalleltransformer} demonstrates that attention outputs for the same query and different keys/values can be composed by preserving both the attention outputs and their scales. Let \(\mathbf{q}\) be a query, and let \(\mathcal{I}\) be an index set. We define the \emph{attention scale} over \(\mathcal{I}\) via the log-sum-exp operation on the attention scores:

\begin{equation}
    \mathbf{LSE}(\mathcal{I}) = \log \sum_{i \in \mathcal{I}} \exp(\mathbf{q} \cdot \mathbf{k}_i)
\end{equation}

where \(\mathbf{k}_i\) is the \(i\)-th key vector. The corresponding \emph{attention output} \(\mathbf{O}(\mathcal{I})\) is then

\begin{equation}
    \mathbf{O}(\mathcal{I}) = \sum_{i \in \mathcal{I}} \frac{\exp(\mathbf{q} \cdot \mathbf{k}_i)}{\exp(\mathbf{LSE}(\mathcal{I}))} \cdot \mathbf{v}_i
\end{equation}

We define the \emph{Attention State} for \(\mathcal{I}\) as the tuple of \emph{attention output} and \emph{attention scale}:
\(\begin{bmatrix}\mathbf{O}(\mathcal{I}) \\ \mathbf{LSE}(\mathcal{I}) \end{bmatrix}\). Crucially, the Attention State of \(\mathcal{I}\cup\mathcal{J}\) can be derived by composing the states of \(\mathcal{I}\) and \(\mathcal{J}\). Specifically, introducing an operator \(\oplus\):

\vspace{-1em}
\begin{eqnarray*}
   \begin{bmatrix} \mathbf{O}(\mathcal{I \cup J}) \\ \mathbf{LSE}(\mathcal{I \cup J}) \end{bmatrix} & \!\!\!\!\!\!=\!\!\!\!\!\! &
    \begin{bmatrix} \mathbf{O}(\mathcal{I}) \\ \mathbf{LSE}(\mathcal{I}) \end{bmatrix} \oplus \begin{bmatrix} \mathbf{O}(\mathcal{J}) \\ \mathbf{LSE}(\mathcal{J}) \end{bmatrix} \\ \ & \!\!\!\!\!\! = \!\!\!\!\!\! & 
    \begin{bmatrix} 
        \frac{\exp(\mathbf{LSE}(\mathcal{I})) \mathbf{O}(\mathcal{I}) + \exp(\mathbf{LSE}(\mathcal{J})) \mathbf{O}(\mathcal{J})}{\exp(\mathbf{LSE}(\mathcal{I})) + \exp(\mathbf{LSE}(\mathcal{J}))}
        \\ \log(\exp(\mathbf{LSE}(\mathcal{I})) + \exp(\mathbf{LSE}(\mathcal{J}))) \end{bmatrix}
\end{eqnarray*}
\vspace{-1em}

Since \(\oplus\) is associative and commutative, multiple sets of attention states can be composed in any order. Ring-Attention~\cite{liu2023ring} and Flash-Decoding~\cite{dao2023flash-decoding} utilize this property to offload partial-attention computations, thereby reducing memory usage and improving hardware efficiency. In \sys, the \emph{Attention State} is adopted as the canonical output of an attention operation, and \(\oplus\) serves as the standard reduction operator (analogous to summation in GEMM) on these states.

\subsection{Block/Vector Sparsity}

\label{sec:block-sparse}

Block Compressed Sparse Row (BSR) is a hardware-efficient sparse format that groups non-zero elements into contiguous matrices of size $(b_r, b_c)$, as opposed to the random scattering found in unstructured sparsity. This format offers several advantages over the standard Compressed Sparse Row (CSR) format. BSR improves register reuse efficiency~\cite{im2004blocksparsity, buluc2009bsr} and demonstrates better compatibility with hardware matrix multiplication units on GPUs and NPUs~\cite{narang2017bsr-rnn,gray2017gpu-bsr}. In addition, it provides the ability to skip empty blocks, reducing computational overhead. BSR's efficiency is particularly evident when subcomputations are aligned with hardware matrix multiplication instructions, such as NVIDIA's \texttt{mma} instructions.
Traditionally, tensor core instructions operate on minimal dimensions of 16 (or larger for newer GPUs), leading most block-sparse kernels to use block sizes that are multiples of $(16, 16)$. However, this approach is not always optimal for applications with fine-grained sparsity patterns~\cite{wang2023tcgnn}. Many attention libraries restrict their block sizes to multiples of $(128, 128)$ for block-sparse attention kernels.

Recent research~\cite{chen2021tensorcoressparsity, li2022magicube} has demonstrated that efficient utilization of the tensor core can be achieved with smaller block sizes, such as $(16, 1)$ for matrix $B$ in GEMM, or $(1, 16)$ for matrix $A$ (also known as vector-sparse). This is accomplished by first gathering rows/columns into contiguous shared memory and then applying dense tensor cores to these contiguous shared-memory data. This approach is particularly beneficial for applications with fine-grained sparsity patterns.
\sys builds upon these techniques to support blocks with arbitrary column sizes $B_c$, offering greater flexibility and efficiency in handling diverse sparsity patterns.

\section{Design}
\label{sec:design}

In this section, we introduce the system design of \sys. We begin by presenting the data structure employed in \sys and demonstrate how Block-Sparse Row (BSR) acts as a versatile abstraction for KV cache storage in attention kernels. Next, we discuss the \sys compiler, which supports various attention variants, alongside a dynamic-aware runtime scheduler that facilitates load-balanced scheduling of attention kernels. Finally, we describe the user-level API designed for integrating \sys with existing LLM serving systems.

\subsection{KV-Cache Storage}

\subsubsection{Block-Sparse Matrix as Unified Format}

\begin{figure}[ht]
    \centering
    \includegraphics[width=0.45\textwidth]{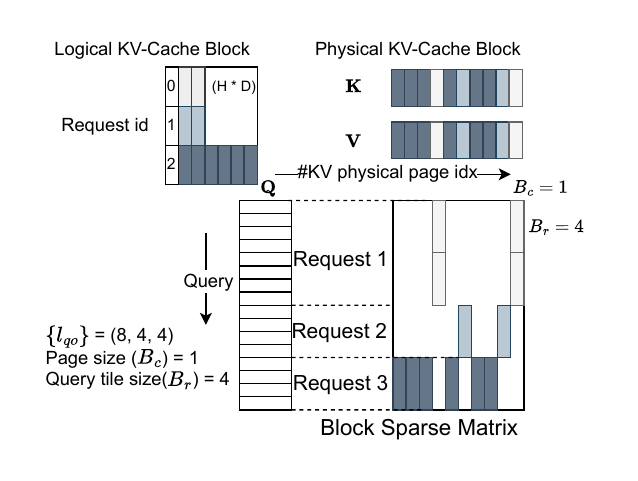}
    \caption{Representation of Page Table in BSR $(B_r=4,B_c=1)$ format. The number of column blocks in the block sparse matrix corresponds to the total number of blocks allocated for the Page Table. Non-zero blocks represent KV-Cache pages accessed by queries.}
    \label{fig:focus-sparse-layout}
\end{figure}

Recent advancements in KV-Cache storage, such as PageAttention~\cite{kwon2023vllm} and RadixAttention~\cite{zheng2023sglang}, employ non-contiguous memory storage with a minimum granularity of a block (or token) of $(H, D)$ tensors, where $H$ represents the number of heads and $D$ the hidden dimension. These structures are optimized to minimize memory fragmentation while enhancing memory reuse and cache hit rates. We demonstrate that these diverse data structures can be unified under a block sparse format, as illustrated in Figure~\ref{fig:focus-sparse-layout}.

The conceptual equivalence between page tables and sparse matrices has been previously explored in SPGrid~\cite{setaluri2014spgrid}, which leverages Translation Lookaside Buffer (TLB) hardware for efficient sparse structure indexing. Beyond page tables and radix trees, sparse matrices can also effectively represent various attention mechanisms, such as Tree Attentions used in speculative decoding~\cite{cai2024medusa,miao2023specinfer,chen2024sequoia} and importance masks applied to KV-Cache~\cite{tang2024quest}.

In \sys, we implement a unified strategy for data representation. Query and output matrices are efficiently stored as ragged tensors (also known as jagged arrays)~\cite{ragged-tensor} without padding, which facilitates the compact packing of queries and outputs from diverse requests into a single tensor. Initially, keys and values are maintained in ragged tensors using the same index pointers as queries, as they originate from the projection matrices $W_q, W_k, W_v$ applied to the same input. These keys and values are subsequently incorporated into the KV-Cache with newly updated entries. The KV-Cache employs a block-sparse row (BSR) format, where block sizes are defined by application requirements: $B_r$ corresponds to the query tile size, details of which will be discussed in later sections, and $B_c$ is specified by KV-Cache management algorithms. \sys kernel implementations supports arbitrary $(B_r, B_c)$ values.

\begin{figure}[ht]
    \centering
    \includegraphics[width=0.45\textwidth]{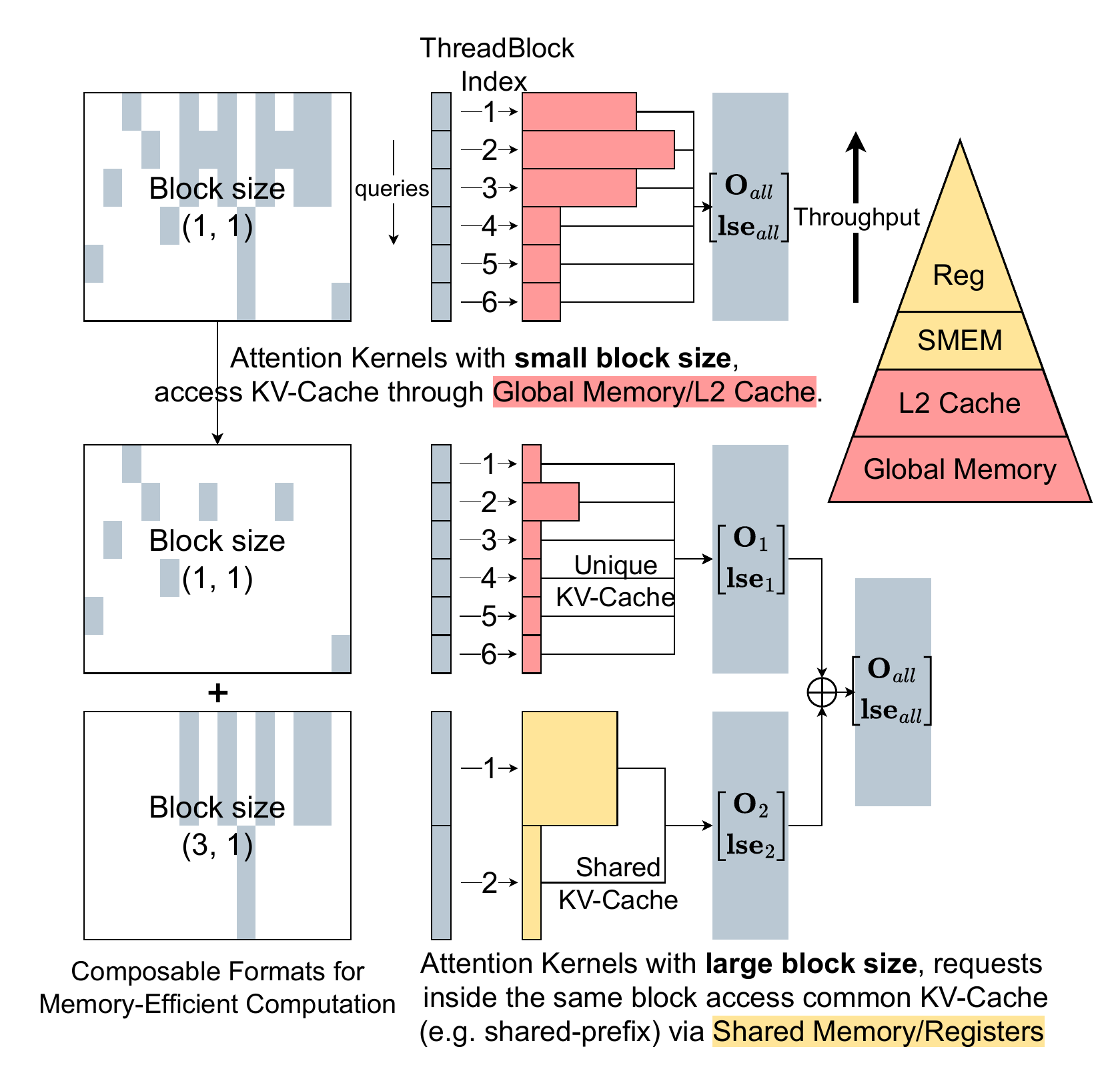}
    \caption{Composable formats for shared-prefix decomposition in attention computation. The queries corresponding to the first 6 rows have a shared prefix, as do the queries in the last 6 rows. We store the KV cache corresponding to the shared prefix in a block sparse matrix with a block size of $(3,1)$, while storing the remaining unique KV cache in another block sparse matrix with a block size of $(1,1)$. For block size $(3,1)$, 3 queries can share the same KV cache in high-bandwidth shared memory, while for block size $(1,1)$, each query access KV-Cache within its own threadblock, which can only go through low-bandwidth global memory or L2 cache.}
    \label{fig:flashinfer-composable-formats}
\end{figure}
\subsubsection{Composable Formats for Memory Efficiency}
\label{sec:composable-formats}

Inspired by SparseTIR~\cite{ye2023sparsetir}, we enhance attention computation efficiency through composable formats. This approach leverages multiple block sparse formats instead of a single format to store the sparse matrix, offering greater flexibility and memory efficiency. Single block-sparse formats are constrained by a fixed block size, limiting memory efficiency based on the number of rows in the block ($B_r$). While larger $B_r$ values improve shared memory and register reuse for requests within the same block, they also increase fragmentation. Conversely, requests in different blocks cannot access each other's shared memory.

Our composable format design allows for the decomposition of the KV cache sparse matrix based on prior knowledge. For instance, if certain requests share a prefix, the corresponding rows and columns in the KV cache form a dense submatrix. We can then use a block sparse matrix with a larger $B_r$ to store these submatrices efficiently. Figure~\ref{fig:flashinfer-composable-formats} illustrates this concept, showing how shared prefixes can be optimized using composable formats. This approach doesn't require data movement in the KV cache; instead, we compute the indices and index pointer arrays for the sparse submatrices. Attention computations on larger block sizes can access shared KV cache entries using fast shared memory and registers, significantly improving memory efficiency.

\subsection{Compute Abstraction}
\label{sec:compute-abs}

We developed CUDA/CUTLASS \cite{cutlass} templates for FlashAttention, designed specifically for both dense and block-sparse matrices and compatible with NVIDIA GPU architectures from Turing to Hopper (sm75 to sm90a). Our implementations utilize the FlashAttention2 (FA2 for short) algorithm \cite{dao2023flashattention2} for architectures up to Ada(sm89), and the FlashAttention3 (FA3 for short) algorithm \cite{jay2024flashattention3} for Hopper. Key improvements include enhanced loading of sparse tiles into shared memory, expanded tile-size configurations, optimized memory access patterns for grouped query attention, and customizable attention variants.

\subsubsection{Global to Shared Memory Data Movement}
\label{sec:sparse-loading}

\begin{figure}[ht]
    \centering
    \includegraphics[width=0.45\textwidth]{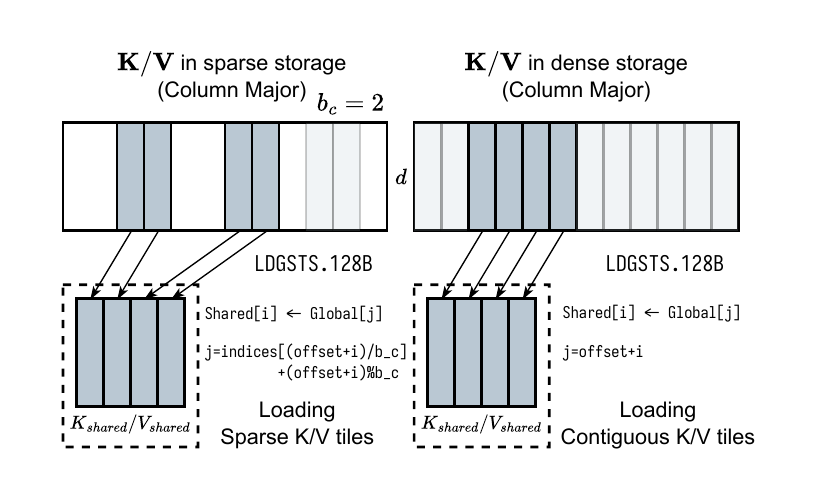} 
    \caption{Data transfer from global to shared memory for sparse/dense KV-Cache in \sys. Left: Sparse KV-Cache with $b_c=2$; Right: Dense KV-Cache. Head dimension $d$.}
    \label{fig:sparse-loading}
\end{figure}

The \sys attention template supports any block size, requiring a specialized data loading approach since blocks may not align with tensor core shapes. As discussed in Section \ref{sec:block-sparse}, we address these challenges by transferring tiles from scattered global memory to contiguous shared memory for dense tensor core operations. Tensor core inputs for a single MMA instruction can originate from different blocks within a block-sparse matrix. Figure \ref{fig:sparse-loading} illustrates how \sys loads tiles from sparse/dense KV-Cache into shared memory; sparse KV-Cache addresses are computed using the \texttt{indices} arrays of the BSR matrix, while dense ones use row index affine transformations.

The last dimension of the KV-Cache remains contiguous (with size of head dimension $d$, commonly 128 or 256), maintaining coalesced memory access that fits GPU cache line sizes. We use asynchronous copy instructions \texttt{LDGSTS} with a 128B width to maximize memory bandwidth. Although the Tensor Memory Accelerator (TMA) in Hopper architecture can further accelerate data movement, it doesn't support non-affine memory access patterns. Consequently, we only use TMA for contiguous KV-Cache on Hopper GPUs and fall back to Ampere-style asynchronous copies for other settings where TMA isn't suitable.

Post-transfer to shared memory, the sparse and dense FlashAttention implementations converge, allowing consistent kernel usage with variations only in data loading modules.

\subsubsection{Microkernel with Different Tile Sizes}
\label{sec:tile-size-selection}

To adapt to the varying operational intensities of LLM applications, \sys implements the FA2 algorithm across multiple sizes. Traditional FA2 uses limited number of tile sizes (e.g., $(128,64)$), optimal for prefill on A100 but inefficient for shorter-query-length decoding. One architecture's ideal tile size may not suit others; for instance, Ada(sm89) has limited shared memory, affecting SM occupancy with large tiles.

\sys offers FA2 kernels with tile sizes ${(1, 16, 32, 64, 128) \times (32, 64, 128)}$ and selects using heuristics based on hardware resources and workload intensity:

\begin{enumerate}[itemsep=2pt,topsep=0pt,parsep=0pt]
\item Determine average query length (for Grouped-Query Attention, the query length are fused with head group dimension, see Appendix \ref{sec:head-group-fusion}) per batch, choosing the minimal query tile size meeting or exceeding it.
\item Formulate register and shared memory constraints as functions of K/V tile size, maximizing SM resource occupancy.
\end{enumerate}

\begin{figure*}[!ht]
    \centering
    \includegraphics[width=1\textwidth]{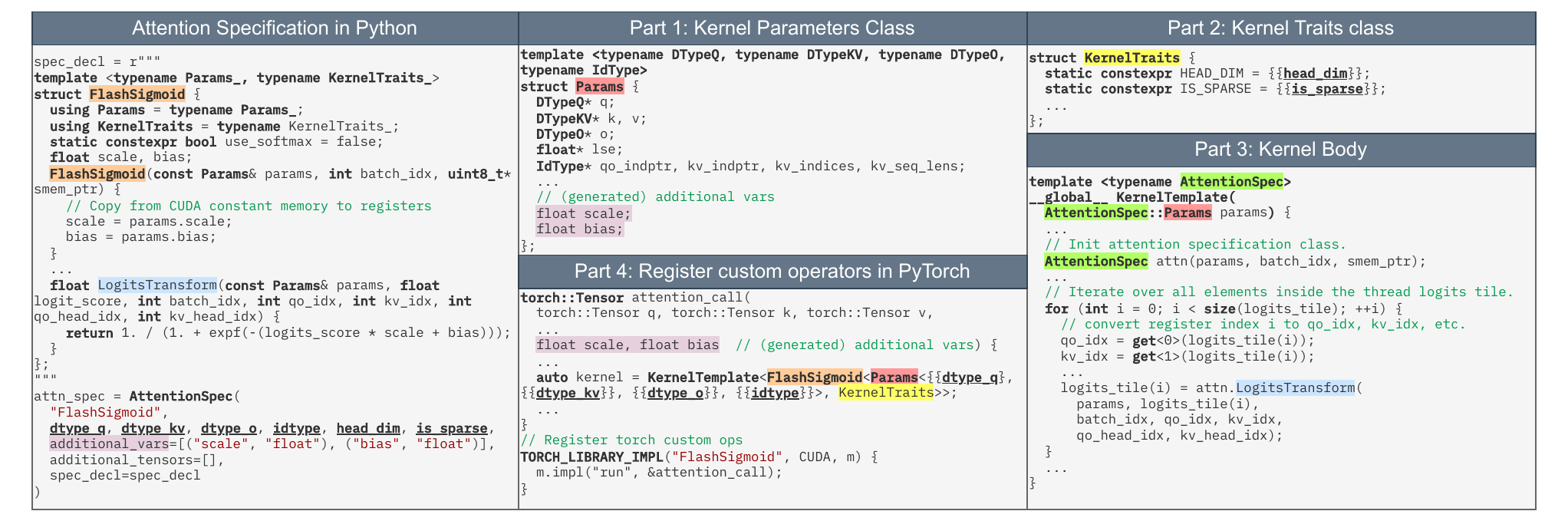}
    \caption{JIT compiler for attention variants in \sys, featuring CUDA code strings defining variant functors, additional variables/tensors, and data types, used to populate kernel templates. Corresponding codes share highlighting.}
    \label{fig:flashinfer-attention-variants}
\end{figure*}
\subsubsection{JIT Compiler for Attention Variants}

For query tile size $1$, we use CUDA Cores template since tensor core instruction $m$ (minimum rows) is $16$, and use Tensor Cores for other query tile sizes. For FA3 we provides row tile sizes that are multiples of $64$, aligning with  Hopper's WGMMA requirements. Tile sizes resolve at compile-time considering task specifics (decoding, prefill, etc.) and hardware capabilities. The block row size $B_r$ is block-sparse matrix is aligned with the query tile size $T_q$.

Recent LLM models use variants to standard attention algorithms. Supporting various attention variants in CUDA library is not sustainable because the we specialize the kernel for each variant for maximum performance, and the number of variants is growing rapidly. However, most attention variants have similar structure to the vanilla attention so we can use the same skeleton of FlashAttention kernels with small modifications. Inspired by FlexAttention ~\cite{he2024flexattention}, we designed a customizable CUDA template and a JIT compiler that takes the attention variant specification as input and generates the optimized kernel code. The variant specification includes the following functors:

\begin{itemize}[itemsep=2pt,topsep=0pt,parsep=0pt]
    \item \texttt{QueryTransform, KeyTransform, ValueTransform}: The transformation applied to the query/key/value tensor before the attention computation.
    \item \texttt{OutputTransform}: The transformation applied to the attention output tensor before returning.
    \item \texttt{LogitsTransform, LogitsMask}: The transformation applied to the logits tensor before the softmax computation, and the mask applied to the logits tensor.
\end{itemize}

Each functor has a fixed signature that takes the kernel parameters, input and current query/key/head index as input, and returns the output. Those variant functions are specified as member of a user-defined variant class which creates a closure for the variant functors. Functors such as \texttt{LogitsTransform} and \texttt{LogitsMask} are inspired by FlexAttention ~\cite{he2024flexattention} and can be used to implement the attention variants with customized logits postprocessing such as custom mask, logits soft-cap ~\cite{morgane2024gemma2,grok} and sliding window attention ~\cite{beltagy020longformer}. \sys has an option of using softmax or not in the attention variant specification, which makes it capable of supporting attention variants that don't use softmax, such as FlashSigmoid ~\cite{jason2024flashsigmoid}. \sys's query and key transformation functors making it possible to fuse normalization, RoPE ~\cite{su2024rope} and projection ~\cite{deepseekv2} into the attention kernel.

Figure \ref{fig:flashinfer-attention-variants} shows how \sys maps FlashSigmoid's ~\cite{jason2024flashsigmoid} attention specification to different parts of \sys's CUDA templates. Attention specificiation accepts a piece of CUDA code to define the variant functors, such design also enables user to use advanced PTX instructions \footnote{\url{https://docs.nvidia.com/cuda/parallel-thread-execution/}} or even their own libraries. The JIT compiler generates the CUDA code by inserting the variant class and other information into the template, and the generated CUDA code is compiled with PyTorch's JIT compiler \footnote{\url{https://pytorch.org/tutorials/advanced/cpp_extension.html\#jit-compiling-extensions}} and registered as a custom operator \footnote{\url{https://pytorch.org/tutorials/advanced/custom_ops_landing_page.html}}. We also support compiling to other runtime systems through a framework agnostic DLPack \footnote{\url{https://github.com/dmlc/dlpack}} interface.

\subsection{Dynamism-Aware Runtime}

In this section we introduce the runtime design of \sys, including the dynamic scheduling framework, and the composable formats for memory efficient attention.

\subsubsection{Load-balanced Scheduling}

\label{sec:load-balanced-scheduling}

In \sys, the load-balanced scheduling algorithm aims to minimize SM idle time by distributing the workload evenly across all SMs. It takes the sequence length information of the query/output and key/value dimensions as input, and produces both the mapping between the workload and Cooperative Thread Arrays (CTAs) and the index mapping for partial and final outputs. The algorithm is presented in Algorithm~\ref{alg:load-balancing} (the head dimension is omitted for simplicity). Our approach is inspired by Stream-K~\cite{osama2023streamk}; however, because LLM serving requires deterministic outputs, we did not incorporate atomic aggregation in Stream-K implementation to avoid non-deterministic behavior. The scheduling algorithm generates deterministic aggregation order when provided with identical sequence length information.
\begin{algorithm}[tb]
   \scriptsize
   \caption{\sys's balanced scheduling algorithm}
   \label{alg:load-balancing}
\begin{algorithmic}[1]
   \STATE {\bfseries Input:} $\{l_{qo}(i), l_{kv}(i)\}_i$, query tile size $T_q$.
   \STATE Define the cost of a tile $l_q, l_{kv}$ as ($\alpha, \beta$ are hyperparameters):
   $$\textrm{cost}(l_q, l_{kv}) = \alpha l_q + \beta l_{kv} $$
   \STATE Compute the maximum KV chunk size $L_{kv}$ by
   \vspace{-1em}
   $$L_{kv} \leftarrow \frac{\sum_{i}\ceil{\frac{l_{qo}(i)}{T_q}}\cdot l_{kv}(i)}{\#\textrm{CTA}}$$
   \STATE Split each query tile's KV into chunks, with maximum size $L_{kv}$, we assign each chunk a work index $w$, and the length of the chunk is $l_{kv}(w)$.
   \STATE Let $W = \{(w, l_{kv}(w))\}$ and sort the entries in descending order of length.
   \STATE $Q \leftarrow \textrm{PriorityQueue}(\{(c, 0)\})$ where $c$ is the CTA index.
   \WHILE{$W \neq \emptyset$}
      \STATE $c, \textrm{current\_cost} \leftarrow Q.\textrm{pop}_{min}()$
      \STATE $w, l_{kv}(w) \leftarrow W.\textrm{pop}()$
      \STATE $\textrm{new\_cost} \leftarrow \textrm{current\_cost} + \textrm{cost}(T_q, l_{kv}(w))$
      \STATE Assign chunk $w$ to CTA $c$
      \STATE $Q.\textrm{push}((c, \textrm{new\_cost}))$
   \ENDWHILE
\end{algorithmic}
   \normalsize
\end{algorithm}

\begin{figure}[ht]
    \centering
    \includegraphics[width=0.4\textwidth]{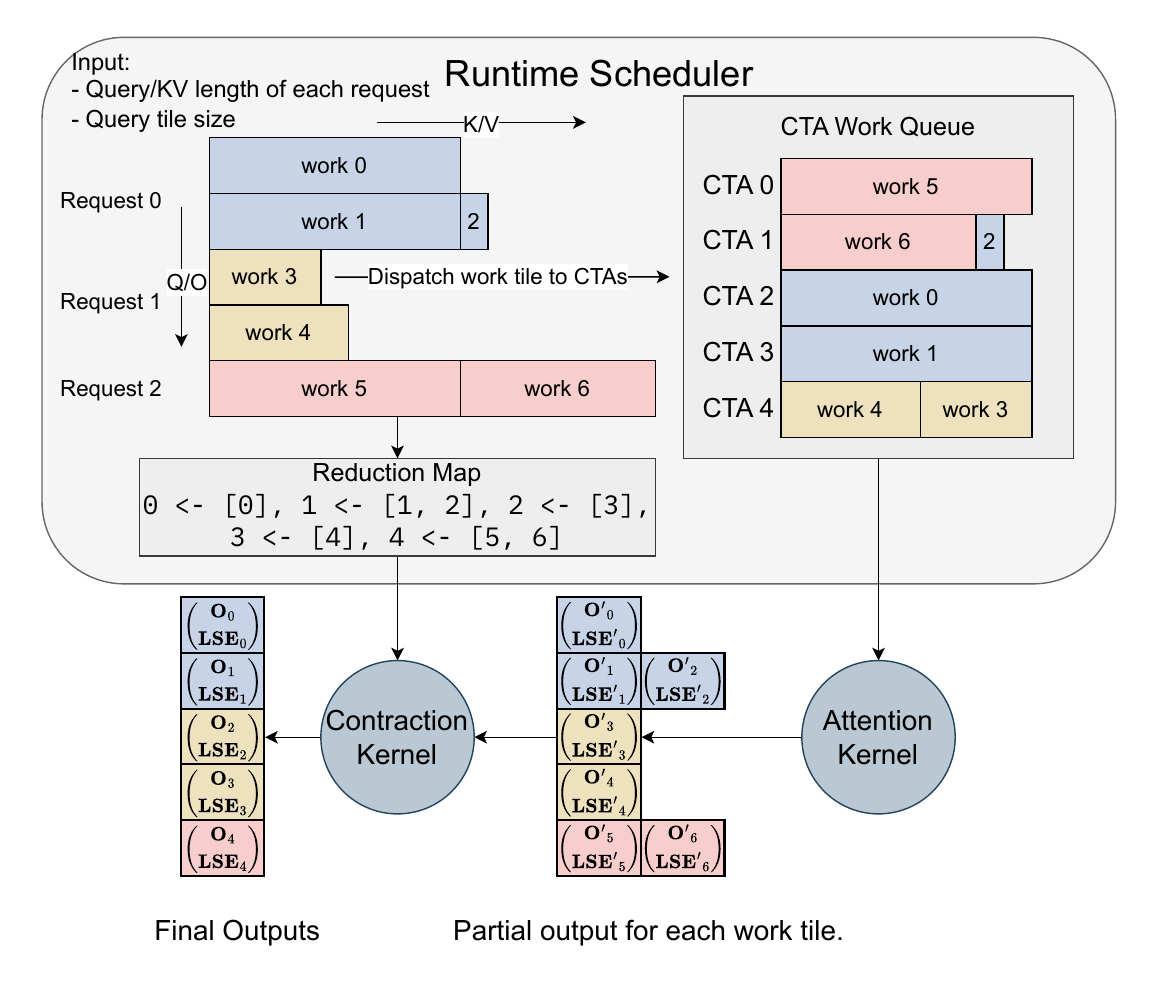}
    \caption{\sys's load-balanced runtime scheduler, sequence length information (both on query/output and key/value dimension) are provided to the scheduler to compute the plan information: (1) Work queue of each CTA (2) Index mapping between partial and final outputs. These plan information are cached at GPU-side and used as inputs for persistent attention/contraction kernels.}
    \centering
    \label{fig:flashinfer-scheduler}
\end{figure}

Figure \ref{fig:flashinfer-scheduler} shows the workflow of \sys's runtime scheduler. The attention kernel do not produce the final output directly because some long KV are split into multiple chunks, and the final output is the contraction (using the attention composition operator mentioned in section \ref{sec:attention-composition}) of all chunks' partial outputs.
The partial outputs are stored in a workspace buffer provided by the user (see section \ref{sec:programming-interface}).
\sys implements efficient attention composition operator that can deal with variable length aggregation. The work queue of each CTA, and the mapping between partial and final outputs need to be planned by the scheduler. Once plan information is computed on CPU, \sys asynchronously copy the plan information to a specific region of the workspace buffer on GPU, and the plan information is used as inputs for persistent attention/contraction kernels. The scheduler runs per generation step to produce plan information as the sequence length changes for each generation step on CPU, and overhead can be amortized over multiple layers because the same plan information can be reused for all layers.

\sys guarantees both attention and contraction stage are compatible with CUDAGraphs ~\cite{alan2019cudagraph,vinh2021pytorchcudagraph}. Both attention and contraction stage use persistent kernel and the grid size is fixed once compiled, which means the kernel is launched with the same grid size for each generation step. We set the fixed offset for each section of the workspace buffer to store partial outputs and plan information to make sure the pointers passed to the kernel are the same for each generation step, meeting the requirement of CUDAGraphs (see Appendix \ref{sec:cudagraph-compatible-layout} for details). We merge the two stages into one persistent kernel, eliminating intra-kernel overhead.

\subsection{Programming Interface}
\label{sec:programming-interface}

\sys offers a programming interface designed for seamless integration with existing LLM serving frameworks such as vLLM\cite{kwon2023vllm}, MLC-Engine\cite{mlcllm}, and SGLang~\cite{zheng2023sglang}.

\begin{lstlisting}[style=zihao-style, morekeywords={with}, label={lst:programming-interface}, caption={\sys PyTorch Programming Interface}]
# Create workspace buffer
workspace = torch.empty(...)
seqlen_info.init()

# Compile: create CUDAGraphs
graphs = []
for task_info in task_infos:
  # Init: compile kernels according to spec
  attn = AttentionWrapper(attn_spec, task_info, workspace)
  g = torch.cuda.CUDAGraph()
  # Dummy plan
  attn.plan(seqlen_info)
  # Capture CUDA graphs
  with torch.cuda.graph(g):
    for i, layer in enumerate(layers):
      ...
      attn.run(...)
      ...
  graphs.append(g)

# Runtime: select the best CUDAGraph
g = select_graph(graphs)
finished = False
# Text generation loop
while not finished:
  seqlen_info.update()
  # Plan per generation step
  attn.plan(seqlen_info)
  # Replay CUDA-Graph
  g.replay()
\end{lstlisting}

Listing \ref{lst:programming-interface} shows the PyTorch programming interface of \sys. The user initializes the wrapper by providing the attention variant specification, task information, and a user-allocated workspace buffer (see Appendix \ref{sec:memory-mgmt} for details) to store partial output and plan information for \sys dynamic scheduling.
Kernel are JIT-compiled at init time and cached for reuse. For composable formats (section \ref{sec:composable-formats}), \sys creates multiple attention wrappers, each with distinct block sizes. Kernels with different average query length and composable format configurations are compiled and captured in different CUDAGraphs. At runtime, the serving framework selects the most appropriate CUDAGraph based on the current KV-Cache configuration, ensuring optimal performance for varying workload characteristics.

The \texttt{plan} function activates the dynamic scheduler by processing sequence length data to generate load-balanced scheduling plans. These plans are cacheable, allowing reuse across operators with matching sequence length specs, such as all decode attentions in a generation step. The \texttt{run} function executes the attention computation using inputs of query, key, value, and cached plan data, outputting the attention results. 
CUDAGraph can capture calls to \texttt{run} functions and compile the entire attention generation step into a single graph. However, \texttt{plan} function is not captured by CUDAGraph because it's on CPU. The \texttt{plan} and \texttt{run} division is inspired by the Inspector-Executor (IE) model \cite{ravi1988principles, saltz1991preprocessed, ravi1993runtime},  which is widely used for parallelizing irregular workloads.

\section{Evaluation}
\label{sec:evaluation}

In this section, we evaluate \sys v0.2 on kernel-level and end-to-end performance showing how \sys's design address the challenges of LLM serving. We achieve 29-69\% inter-token-latency reduction compared to Triton backend for LLM serving benchmark, 28-30\% latency reduction for long-context inference, and 13-17\% speedup for LLM serving with parallel generation. We conduct experiments on NVIDIA A100 40GB SXM and H100 80GB SXM GPUs, using CUDA 12.4 and PyTorch 2.4.0 and f16 precision for storage and computation.

\subsection{End-to-end LLM serving performance}

We evaluate \sys with SGLang v0.3.4 \cite{zheng2023sglang} and compare its performance against two settings: SGLang with Triton v3.0 \cite{triton}. The latter is a leading LLM serving engine optimized for NVIDIA GPUs; however, its attention kernels are closed-source, which limits transparency and potential for community-driven improvements. To ensure a comprehensive evaluation, we employ two datasets: the widely-used ShareGPT dataset \footnote{\url{https://huggingface.co/datasets/anon8231489123/ShareGPT_Vicuna_unfiltered/resolve/main/ShareGPT_V3_unfiltered_cleaned_split.json}} and a synthetic workload (Variable) with sequence lengths uniformly distributed between 512 and 2048 tokens. We measure the TTFT(time-to-first-token) and ITL(inter-token-latency) under latency-sensitive online serving settings, the request rate is adjusted to maintain P99 TTFT below 200ms.

\begin{figure}[ht]
    \centering
    \includegraphics[width=0.45\textwidth]{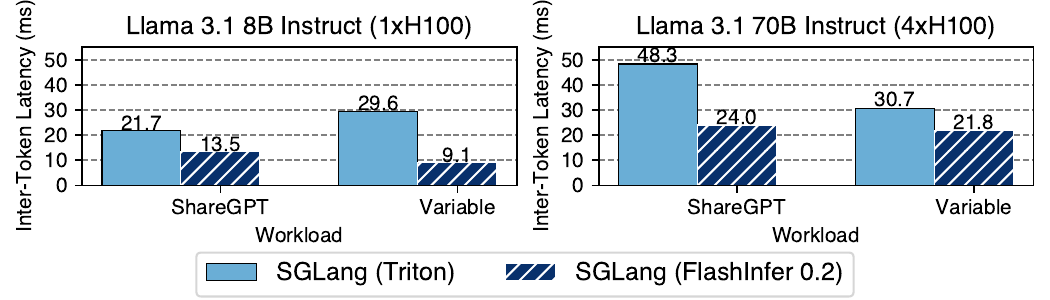}
    \includegraphics[width=0.45\textwidth]{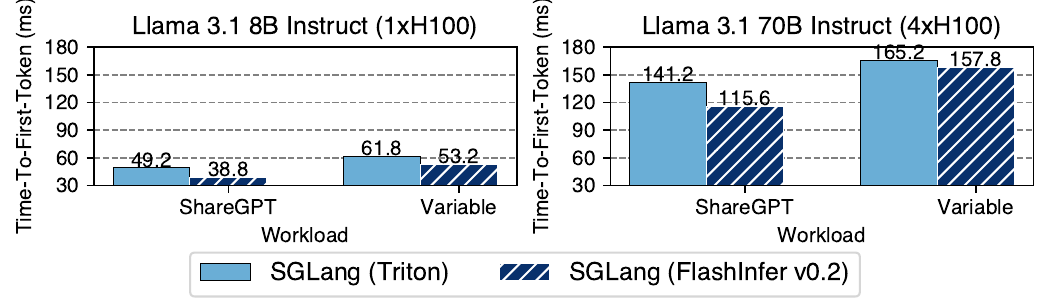}
    \caption{Medium Inter-Token-Latency (ITL) and medium Time-To-First-Token (TTFT) of SGLang integrated with \sys and Triton.}
    \label{fig:sglang-itl-ttft}
\end{figure}

Figure \ref{fig:sglang-itl-ttft} shows the ITL and TTFT measured on both Llama 3.1 ~\cite{llama31} 8B (on 1xH100) and Llama 3.1 70B (on 4xH100) models. Compared to SGLang with Triton backend, \sys backend shows consistent speedup in all settings.

\subsection{Kernel Performance for Input Dynamism}
\label{eval:input-dynamism}

In this section we measure \sys's generated kernel performance against state-of-the-art open-source FlashAttention library under different sequence length distributions, we use the latest main branch \footnote{Commit: \href{https://github.com/Dao-AILab/flash-attention/commit/c1d146cbd5becd9e33634b1310c2d27a49c7e862}{c1d146c}} which includes both FlashAttention2 and FlashAttention3 kernels. We fix the batch size to 16 and select three different sequence length distributions: constant (1024), uniform (512 to 1024) and skewed (Zipf distribution with average length 1024). For prefill kernels, we enabled causal masking because it's a common setting in LLM serving.

\begin{figure}[ht]
    \centering
    \includegraphics[width=0.45\textwidth]{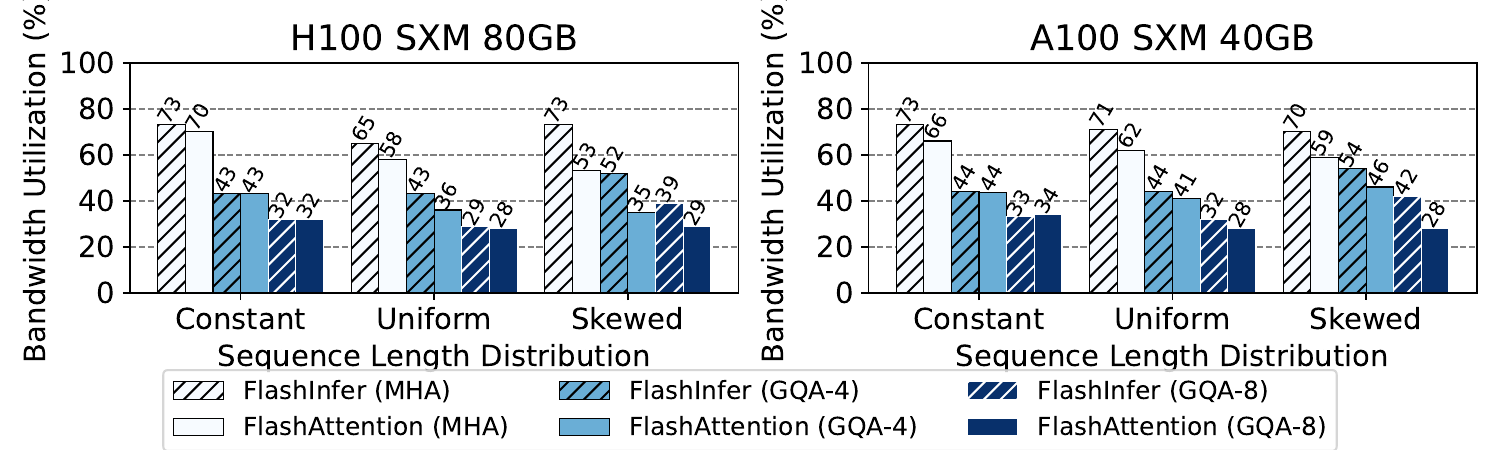}
    \includegraphics[width=0.45\textwidth]{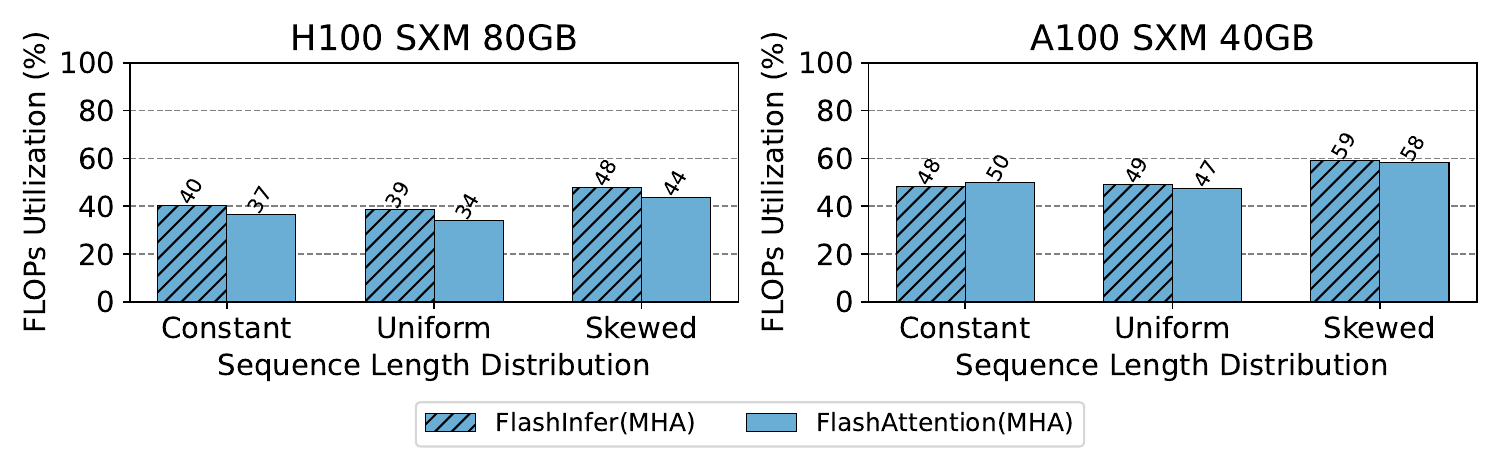}
    \caption{Achieved bandwidth and FLOPs utilizations (the higher the better) for decode (top) and prefill (down) kernels.}
    \label{fig:varlen}
\end{figure}

Figure \ref{fig:varlen} show the achieved bandwidth and FLOPs utilization for decode and prefill kernels. \sys's kernel significant outperforms FlashAttention kernels in uniform and skewed sequence length distributions because of our  load-balanced dynamic scheduler (section \ref{sec:load-balanced-scheduling}). \sys's decode attention outperforms FlashAttention kernels because our versatile tile size selection (section \ref{sec:tile-size-selection}) and FlashAttention use suboptimal tile size for decoding.

\subsection{Customizability for Long-Context Inference}

In this section, we demonstrate how \sys's customized attention kernels significantly accelerate LLM inference. We focus on Streaming-LLM \cite{xiao2023streamingllm}, a recent algorithm capable of million-token inference with constant GPU memory usage. While Streaming-LLM requires specialized attention kernels for optimal performance, particularly a fused kernel combining RoPE \cite{su2024rope} with attention, \sys can generate such fused kernels with merely 20 additional lines of code for query/key transformations. We compare the performance of \sys-generated fused kernels against un-fused kernels (both \sys's and FlashAttention's) and quantify the end-to-end latency reduction achieved by integrating \sys kernels into StreamingLLM.
\begin{figure}[ht]
    \centering
    \includegraphics[width=0.4\textwidth]{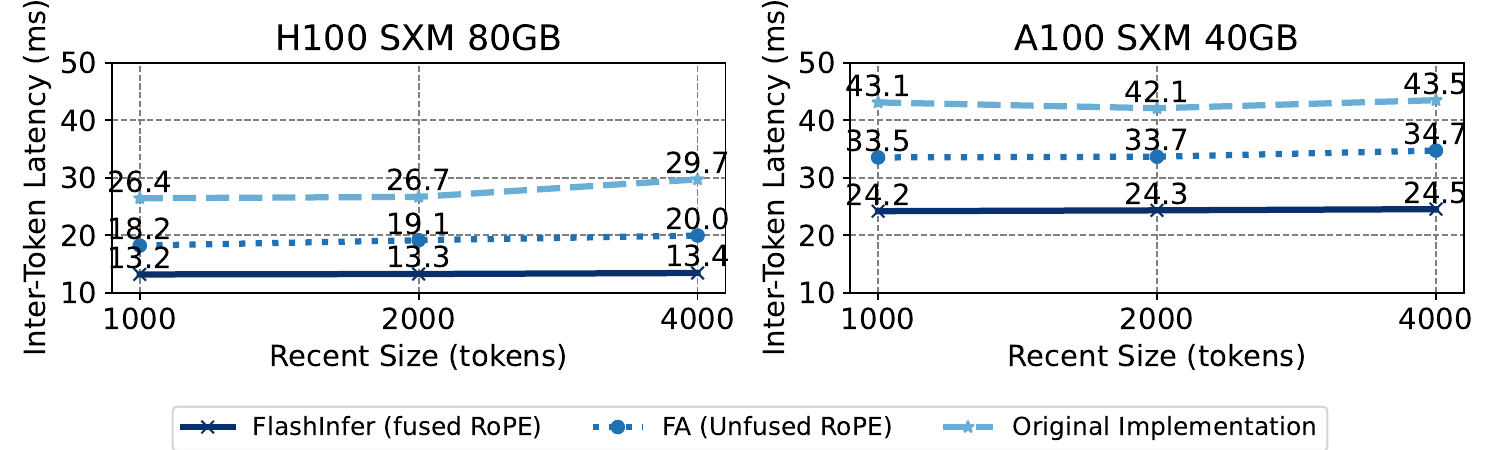}
    \includegraphics[width=0.4\textwidth]{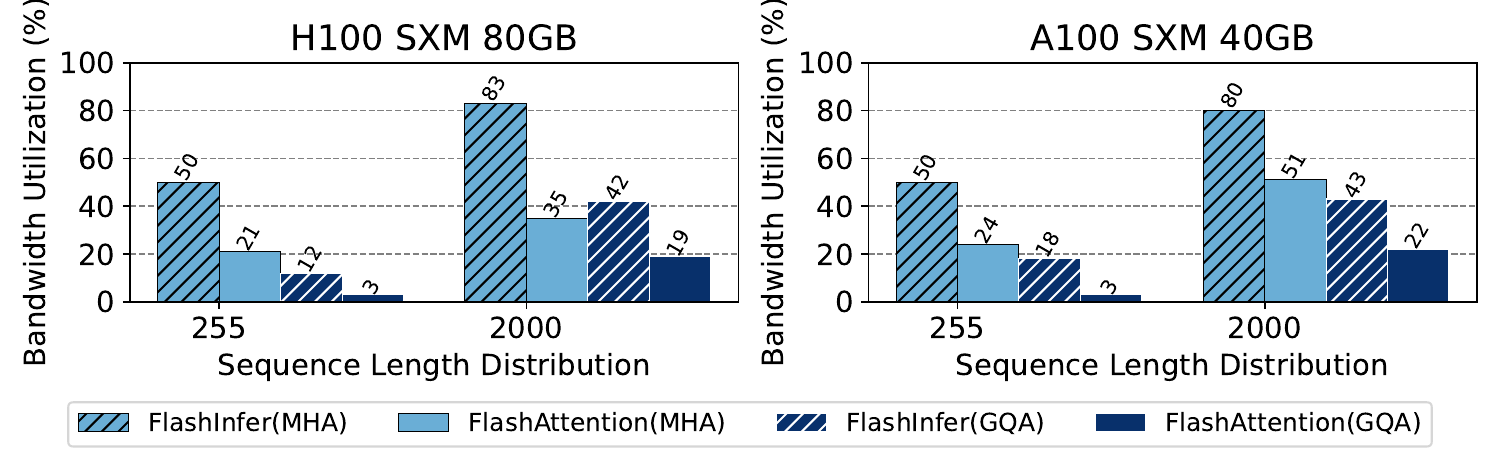}
    \caption{Top: End-to-end latency of Streaming-LLM with \sys fused and FlashAttention's unfused kernels, original implementation is included. Down: bandwidth utilization of \sys fused RoPE kernel compared to FlashAttention's unfused kernel.}
    \label{fig:streaming-llm}
\end{figure}

For end-to-end performance, we run Vicuna-13B ~\cite{vicuna2023} inference on MT-Bench ~\cite{zhenbg2023mtbench} dataset and measure the inter-token-latency (ITL) of Streaming-LLM with and without \sys kernels. Figure \ref{fig:streaming-llm} show the ITL of Streaming-LLM with and without \sys fused  kernels on our optimized implementation of Streaming-LLM (we noticed that the original implementation is sub-optimal and have unnecessary overheads). \sys's fused kernel can yield $28-30\%$ latency reduction under different settings (by changing the recent window size of Streaming-LLM). Original implementation is included as a baseline reference. We also show the kernel-level performance comparison between \sys's fused RoPE kernel and the combination of FlashAttention's RoPE kernel and FlashAttention's attention kernel. \sys's fused RoPE kernel achieves 1.6-3.7x higher bandwidth utilization compared to not fusing attention with RoPE, which necessitate the importance of customizability of attention kernels.

\subsection{Parallel-Generation Performance}
\label{eval:memory-heterogeneity}

In this section, we illustrate how the composable formats of \sys can enhance parallel decoding. With parallel generation emerging as a significant task in LLM serving, it offers great utility in LLM agents. The OpenAI API provides an \texttt{"n"} parameter\footnote{\url{https://platform.openai.com/docs/api-reference/chat/create}} to facilitate the generation of multiple tokens simultaneously. As shared prefixes often exist, prefix-caching can significantly boost the efficiency of parallel generation. The composable formats found in \sys (see Section \ref{sec:composable-formats}) allow for the decoupling of attention computation between the shared prefix and the subsequent suffix, which can be leveraged to expedite parallel decoding.
\begin{figure}[ht]
\centering
\includegraphics[width=0.4\textwidth]{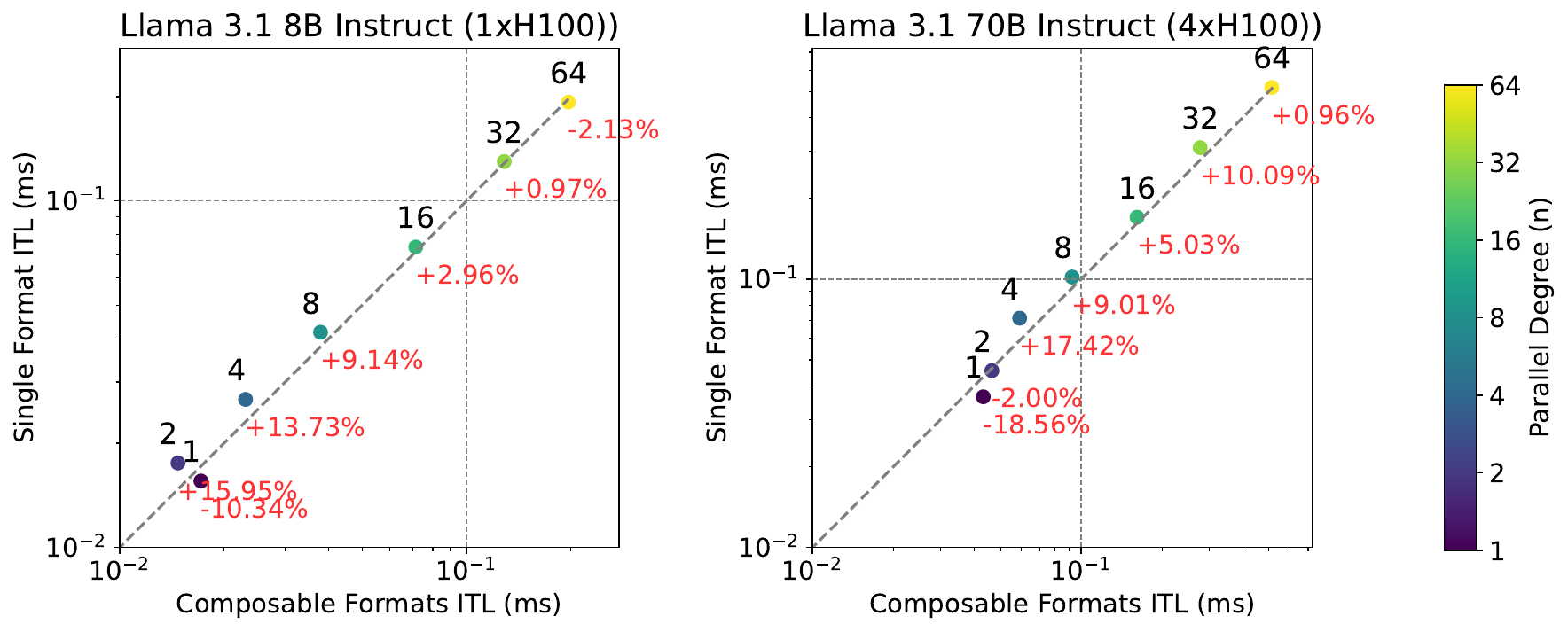}
\includegraphics[width=0.4\textwidth]{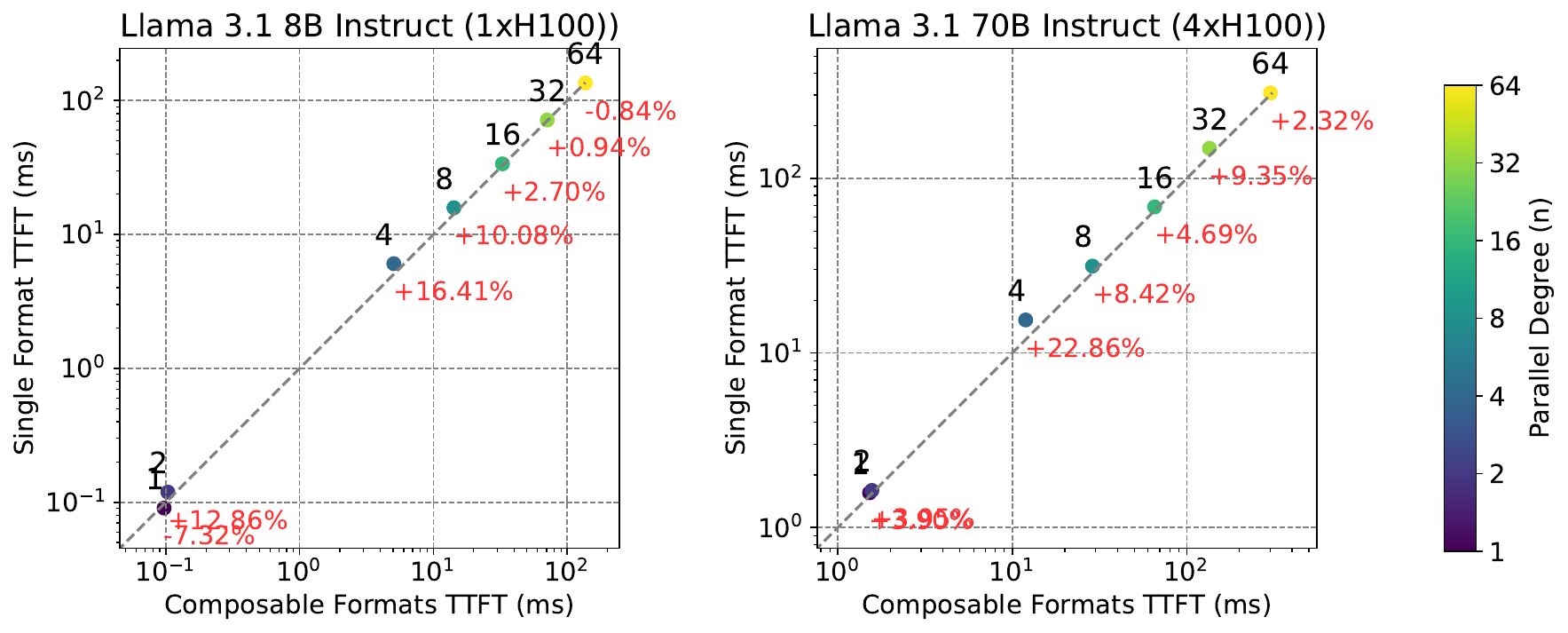}
\caption{ITL and TTFT of MLC-Engine with and without composable formats during parallel generation, the x-axis refers to composable formats performance, and the y-axis refers to single format performance, if a point is above the diagonal line, it means composable formats outperform single format. Different parallel generation $n$ are shown in different colors.}
\label{fig:parallel-generation}
\end{figure}

We implemented composable formats within MLC-Engine\cite{mlcllm} under a prefix-caching configuration and assessed the performance during parallel generation. Evaluations were conducted on the Llama 3.1 models with 8B and 70B parameters\cite{llama31} using the ShareGPT dataset. With a fixed request rate of 16, we varied the number of parallel tokens over the set ${1, 2, 4, 8, 16, 32, 64}$, comparing these results against MLC-Engine configurations where composable formats were disabled. Figure \ref{fig:parallel-generation} presents the ITL (Inference Time Latency) and TTFT (Time To First Token) results for MLC-Engine both with and without composable formats.

For moderate levels of parallel generation ($4 \leq n \leq 32$), \sys's composable formats yield consistent speedups for both ITL and TTFT. Peak speedups occur at $n=4$, where ITL decreases by $13.73\%$ for the 8B model and $17.42\%$ for the 70B model, while TTFT is reduced by $16.41\%$ for the 8B model and $22.86\%$ for the 70B model. Smaller values of $n$ do not benefit from composable formats due to insufficient increase in block size. For larger $n$, the computation ceases to be dominated by attention processes (especially in the case of ShareGPT with its short sequence length), causing the advantage of composable formats to plateau.

\section{Related Work}
\label{sec:related-work}

\subsection{Attention Optimizations}

Multi-Head Attention (MHA)~\citep{vaswani2017transformer} faces computational and IO challenges. FasterTransformer~\citep{fastertransformer} reduces global memory footprint via Fused Multi-Head Attention (FMHA), but doesn't scale to long contexts because shared memory usage is linear to sequence length.
ByteTransformer~\citep{zhai2023bytetransformer} optimizes FMHA on variable-length input.
FlashAttention ~\citep{dao2022flashattention} uses online-softmax ~\citep{milakov2018onlinesoftmax} trick to reduce the shared memory footprint to constant size, enabling long contexts. FlashAttention2\&3~\citep{dao2023flashattention2,jay2024flashattention3} further optimizes FlashAttention by improving loop structure and overlapping softmax and GEMM. FlashDecoding ~\cite{dao2023flash-decoding} applies Split-K to decode attention kernels. LeanAttention~\cite{sanovar2024lean} uses StreamK~\citep{osama2023streamk} to reduce wave-quantization~\citep{wave-quantization} in attetnion (with fixed sequence length). \sys extends the FlashAttention2\&3 template to support sparse attention kernels, while using StreamK-like optimizations on variable length sequences. Nanoflow ~\citep{zhu2024nanoflow} introduces horizontal fusion of GEMM, attention, and communication operations, while POD-Attention ~\citep{kamath2024podattention} focuses on optimizing chunked-prefill attention. The JIT compilation framework of \sys can be extended to generate kernels supporting these fusion techniques. FlashDecoding++~\citep{flashdecodingpp} leverages attention scale statistics to predefine a unified max value. This process converts attention composition (section \ref{sec:attention-composition}) to summation, enabling TMA Store Reduce ~\citep{tma-store-reduce} to asynchronously updating global \textit{attention state}s, it's orthogonal to \sys's contribution and we leave it for future work.

Recent works like RelayAttention~\citep{zhu2024relayattention}, Hydragen~\citep{juravsky2024hydra}, ChunkAttention~\citep{ye2024chunkattention}, and Parrot~\citep{lin2024parrot} explore shared prefix decoding attention but require separate KV-Cache management for prefixes and suffixes. In contrast, \sys's composable formats support multi-level, multiple-prefix decoding with unified page table management, enabling seamless integration into LLM serving frameworks without modifying memory management modules.

\subsection{Sparse Optimizations on GPUs}

FusedMM~\citep{rahman2021fusedmm} explores Sparse-dense Matrix Multiplication (SpMM) fusion, though it omits softmax computation, limiting direct applicability for accelerating attention. \citet{zhang2022fusedgat} explore Graph Attention Networks (GAT) kernel fusion, SAR~\citep{mostafa2022sar} serializes Sparse Attention aggregation, akin to FlashAttention, neither work explores using Tensor Cores. Blocksparse library~\citep{gray2017gpu-bsr} implements BSR GEMM with tensor cores. \citet{chen2021tensorcoressparsity}, TC-GNN~\citep{wang2023tcgnn} and Magicube~\citep{li2022magicube} propose vector sparse formats to leverage Tensor Cores effectively. \sys improves upon these to support any block sizes $(b_r, b_c)$ in FlashAttention.

\subsection{Attention Compilers}

FlexAttention~\citep{he2024flexattention} provides a user-friendly interface for programming attention variants, compiling them into block-sparse flashattention implemented in Triton~\citep{triton}. It uses PyTorch Compiler~\citep{jason2024pytorch2} to automatically generate backward passes. \sys expands the FlexAttention's programming interface to support query/key transformations, and focus on vector-sparsity and load-balancing for LLM serving. \sys generates CUDA code instead of Triton because Triton still underperform CUDA \& CUTLASS in many use cases. \sys can act as a backend for FlexAttention in forward pass. Mirage~\citep{wu2024mirage} optimizes tiling strategies for GEMM and FlashAttention using a probabilistic equivalence verifier, relying on Triton and CUTLASS for code generation. However, it lacks support for variable length and sparse data structures, and doesn't include safe-softmax, unlike \sys, which is directly applicable to LLM serving.

\subsection{LLM Serving Systems}

Orca~\citep{yu2022orca} introduces continuous batching for enhanced throughput. PagedAttention~\citep{kwon2023vllm} uses a Page Table for KV-Cache management. Sarathi-serve~\citep{agrawal2024sarathi} improves efficiency by piggybacking decode operations with chunked-prefill, while SGLang~\citep{zheng2023sglang} utilizes RadixTree for better prefix-caching and KV-management. \sys provides a unified solution for these attention mechanisms through block-sparse attention kernels. vAttention~\citep{prabhu2024vattention} shows that GPU virtual memory can manage address translation in PageAttention without special kernels. Yet, challenges like dynamic KV-Cache sparsity persist, as seen in Quest~\citep{tang2024quest}. Here, \sys's block sparse kernel remains effective. Additionally, \sys can be combined with vAttention by generating kernels for contiguous KV-Cache storage.

\section{Discussions}
\label{sec:discussions}

Currently, \sys only supports the forward pass for attention computation.
To extend \sys and apply it to training would require developing customizable backward attention kernel templates, which we plan to explore in future work.
Regarding the generality of \sys, our approach decouples computation from tile scheduling, allowing for diverse tiling strategies such as FlashDecoding~\cite{flashdecodingpp} and Lean Attention~\cite{sanovar2024lean} by delegating scheduling to a runtime-scheduler rather than embedding it within the attention template.
This design generalizes scheduling algorithms, as detailed in Algorithm \ref{alg:load-balancing}, for load-balancing and wave-quantization reduction while targeting optimal GPU performance across architectures (e.g., FlashAttention2~\cite{dao2023flashattention2} for Turing/Ampere/Ada and FlashAttention3~\cite{jay2024flashattention3} for Hopper).
Although frameworks like Triton~\cite{triton} offer GPU-agnostic interfaces, they often lag in adopting new hardware features; our template design space, expressed as $f_\textrm{epilogue}(\textrm{scan}(f_\textrm{logits}(f_{q}(Q)\cdot f_{k}(K)))\cdot f_{v}(V))$, covers most attention functions, including recent variants such as Multi-head Latent Attention (MLA)~\cite{deepseekv2} and the intra-attention component of Linear Attention~\cite{yang2024gatedlinearattentiontransformers}.

\section{Conclusion and Future Work}
\label{sec:conclusion-future-work}

In this paper, we present \sys, an versatile and efficient attention engine for LLM serving. We propose a unified block-sparse storage and composable formats for memory efficiency, JIT compilation for customization and load-balanced scheduler for input dynamism. We evaluate \sys's performance across diverse inference scenarios, showing strong performance in kernel-level and end-to-end LLM serving metrics. In the future, we plan to explore compiling higher-level DSLs ~\cite{wu2024mirage, he2024flexattention} to attention specifications in \sys, as well as code generation to other backends ~\cite{ozennvdsl, spector2024thunderkittens, triton}. The \sys project is open source and available at \url{https://github.com/flashinfer-ai/flashinfer}, and has been deployed at scale in production-level systems.

\section{acknowledgements}
\label{sec:acknowledgements}

We thank anonymous MLSys reviewers for providing constructive comments, LMSYS ORG, UW Syslab and SAMPL research group, CMU Catalyst group for their useful feedback and discussions, Yaxing Cai, Junru Shao, Lianmin Zheng, Ying Sheng, Liangsheng Yin, Lily Liu, Woosuk Kwon, Cody Yu, Ray Wan, Bowen Wang, Pavani Majety, Elfie Guo, Travis Addair, Cuimi Guo for their help in integrating \sys into LLM serving frameworks, Zhuoming Chen, Lesheng Jin, Antoni Baum, Kaichao You, Simon Mo, Ke Bao, Byron Hsu, Zhiqiang Xie, Haofeng Huang, Sirui Lu, Henry Xiao, Chi-Chih Chang, Yilong Zhao, Size Zheng, Bohan Hou, Yang Yu, Nandor Licker, Tsu Bin, Hieu Pham, Horace He, Vijay Thakkar, Yuxian Qiu, Freddy Qi, June Yang, Bing Xu, Anxhelo Xhebraj, Evghenii Gaburov, Bastian Hagedorn and all community contributors for their input and feedback on the \sys project. This work was supported in part by NSF award CNS-2211882, and gifts from OctoAI, Qualcomm, and CMU opensource software fellowships.
Researchers from UW are supported in part by NSF under award CCF-1518703, and by ACE and PRISM, two of the seven centers in JUMP 2.0, a Semiconductor Research Corporation (SRC) program sponsored by DARPA;
Zihao Ye is supported by the NVIDIA Graduate Fellowship. Luis Ceze is supported by the Lazowska Endowed Professorship. The opinions and conclusions in this paper do not reflect the views of these funding agencies.

\newpage
\bibliography{references}
\bibliographystyle{mlsys2024}

%%%%%%%%%%%%%%%%%%%%%%%%%%%%%%%%%%%%%%%%%%%%%%%%%%%%%%%%%%%%%%%%%%%%%%%%%%%%%%%
%%%%%%%%%%%%%%%%%%%%%%%%%%%%%%%%%%%%%%%%%%%%%%%%%%%%%%%%%%%%%%%%%%%%%%%%%%%%%%%
% SUPPLEMENTAL CONTENT AS APPENDIX AFTER REFERENCES
%%%%%%%%%%%%%%%%%%%%%%%%%%%%%%%%%%%%%%%%%%%%%%%%%%%%%%%%%%%%%%%%%%%%%%%%%%%%%%%
%%%%%%%%%%%%%%%%%%%%%%%%%%%%%%%%%%%%%%%%%%%%%%%%%%%%%%%%%%%%%%%%%%%%%%%%%%%%%%%

\clearpage
\appendix

\section{Head Group Fusion for Grouped-Query Attention}
\label{sec:head-group-fusion}

\begin{figure}[ht]
    \centering
    \includegraphics[width=0.5\textwidth]{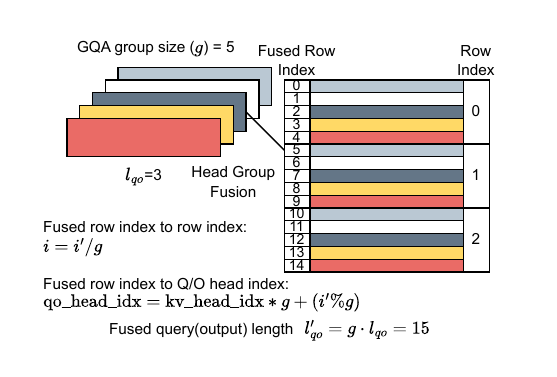}
    \caption{\sys's head-group fusion of query heads with the query length dimension in GQA.}
    \label{fig:flashinfer-gqa-packing}
\end{figure}

Grouped-Query Attention (GQA)~\cite{ainslie2023gqa} allows multiple query heads to share the same key-value (KV) heads. A straightforward implementation that assigns distinct GPU threadblocks to each query head leaves much of the potential KV-Cache reuse underutilized when the query length is short. To address this limitation, \sys offers a \emph{head-group fusion} strategy: 
different KV heads are mapped to individual threadblocks, while query heads are fused with the query length dimension. This fusion scheme is illustrated in Figure~\ref{fig:flashinfer-gqa-packing}, which shows how the fused row index relates to the original row index and the head indices. By merging the query-head dimension with the row dimension in the threadblock mapping, a single shared-memory load of the KV-Cache suffices for all query heads in the group, leading to better memory reuse and improved throughput for GQA operations.

We prefer head-group fusion primarily for short query lengths. When the query length is sufficiently large, the query dimension itself yields enough workload to effectively utilize the KV-Cache, making head-group fusion less critical. Similar ideas have also been explored in other frameworks, such as XQA~\cite{xqa} in TensorRT-LLM~\cite{nvidia2023tensorrt}.

\section{Overhead of Sparse Gathering}
\label{sec:overhead-sparse-gathering}

In Section~\ref{sec:sparse-loading}, we detailed the design of \sys's sparse loading module, which transfers sparse rows from global memory into contiguous shared memory. Here, we measure the performance overhead associated with sparse gathering in \sys for both \emph{decode} and \emph{prefill} kernels.

\begin{figure}[ht]
    \centering
    \includegraphics[width=0.4\textwidth]{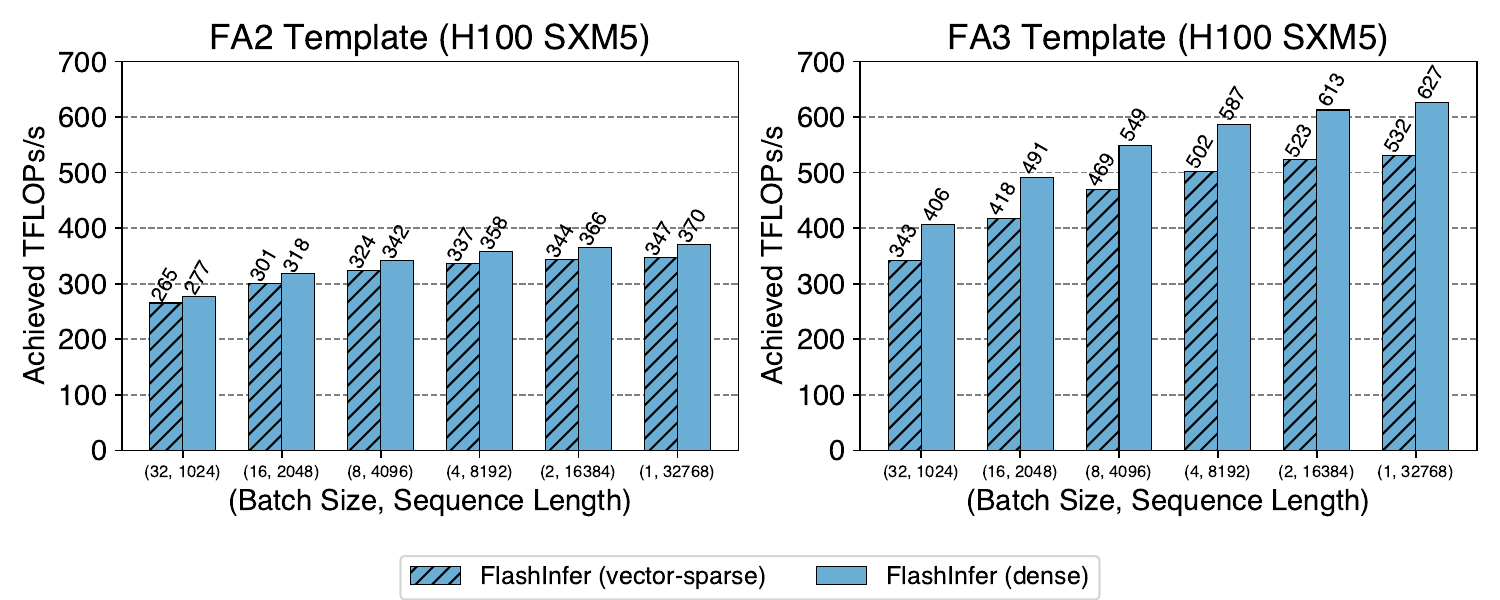}
    \includegraphics[width=0.4\textwidth]{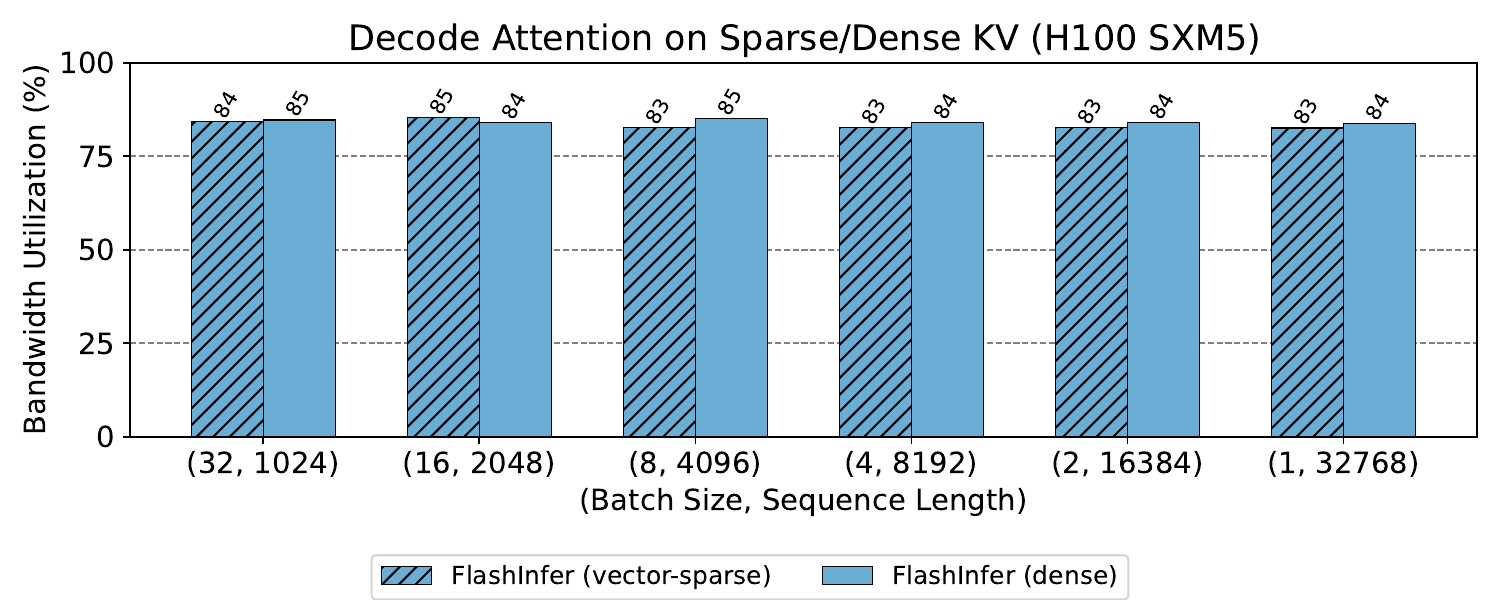}
    \caption{\textbf{Top:} Achieved TFLOPs/s for (causal) prefill attention kernels on FA2/FA3 templates with both dense/sparse KV-Cache. 
    \textbf{Bottom:} Achieved bandwidth utilization for decode attention kernels for both dense/sparse KV-Cache. 
    We use PageAttention with page size 1 (vector-sparse) for sparse KV-Cache. The x-axis shows various batch sizes and sequence lengths.}
    \label{fig:sparse-dense}
\end{figure}

Figure~\ref{fig:sparse-dense} compares achieved throughput in both prefill and decode kernels for sparse and dense (contiguous) KV-Cache.
For the prefill kernels, we measure the \textit{causal} attention scenario, which is common in LLM serving. For contiguous KV-Cache, We use the variable-length RaggedTensor prefill attention API\footnote{\url{https://docs.flashinfer.ai/api/prefill.html\#flashinfer.prefill.BatchPrefillWithRaggedKVCacheWrapper}}. For sparse KV-Cache, we use the PagedKVCache prefill attention API\footnote{\url{https://docs.flashinfer.ai/api/prefill.html\#flashinfer.prefill.BatchPrefillWithPagedKVCacheWrapper}}.

The number of query heads and KV heads are both fixed at 32, head dimension is set to 128. We vary batch size and sequence length to measure the achieved throughput. For decode kernels, the performance gap between sparse and dense KV-Cache is negligible (within $1\%$). For prefill kernels, there is approximately a $10\%$ performance gap.

Note that dense attention in the FA3 template uses TMA instructions (Tensor Memory Access) for key/value loading, which is unavailable for sparse gathering because Hopper Architecture's TMA only supports fixed-stride accesses, whereas sparse gathering requires arbitrary row indices. Consequently, sparse gathering on FA3 relies on Ampere-style asynchronous copy instructions and manual pointer arithmetic. This approach consumes more registers and necessitates smaller KV-tile size to avoid register spilling, leading to a slightly larger performance gap. By contrast, in the FA2 template (where both sparse and dense use Ampere's async-copy), the gap is smaller because the same tile size is used.

When the block column size in a block-sparse matrix is large (e.g., 128 or greater), TMA can be used for sparse gathering since each TMA instruction operates within a single block with fixed stride. We leave this optimization for future work. However, increasing the block column size reduces the flexibility of the block-sparse format, which might not be suitable for all use cases.

\section{The Choice of Backend}
\label{sec:backend-choice}

For NVIDIA GPUs, we build \sys on top of CUDA/CUTLASS~\cite{cutlass} instead of Triton~\cite{triton} for the following reasons:
\begin{enumerate}
    \item \textbf{Advanced NVIDIA GPU Features.} CUTLASS supports specialized GPU capabilities such as warp-specialization~\cite{warp-specialization} and TMA instructions~\cite{hopper-tma}, which are experimental or unsupported in Triton at this moment.
    \item \textbf{Fine-Grained Kernel Optimization.} While Triton provides tile-level abstractions, CUDA/CUTLASS affords finer control over thread-level registers. This flexibility simplifies incorporating low-level optimizations (e.g., PTX intrinsics) directly into our JIT templates, which is more challenging in Triton.
\end{enumerate}

Our load-balancing scheduler design (Section~\ref{sec:load-balanced-scheduling}) is largely backend-agnostic, allowing us to potentially integrate Triton in future versions of \sys and to adapt our approach to other hardware platforms.

\section{Memory Management}
\label{sec:memory-mgmt}

\sys manages a page-locked (pinned) host buffer and a device workspace buffer to store scheduler metadata and split-k partial outputs. We divide the device workspace buffer into \emph{sections}, each corresponding to an array of either scheduler metadata or partial split-k outputs. For each \lstinline{plan} call in the scheduler, we compute the scheduler metadata on the pinned host buffer and then issue a \lstinline{cudaMemcpyAsync} to transfer this data into the corresponding sections of the device workspace buffer.

\subsection{CUDAGraph-Compatible Workspace Layout}
\label{sec:cudagraph-compatible-layout}

Once a kernel is captured by CUDA Graph, its arguments (pointers and scalars) become fixed, implying that each section of the device workspace buffer must maintain a consistent address for the entire captured graph's lifetime. Therefore, we allocate the workspace buffer to its maximum required capacity for each section, based on upper-bound estimations of scheduler metadata and partial outputs.

\subsection{Split-K Writethrough Optimizations}
\label{sec:writethrough-optimizations}

In \sys's load-balancing scheduler (Section~\ref{sec:load-balanced-scheduling}), KV-splitting is only applied to requests that have large KV lengths. Requests with short KV lengths do not require splitting and hence have no reduction step from partial output. To save both computation and workspace memory, these small requests can write their partial outputs directly to the final output buffer (bypassing the device workspace buffer). This approach reduces both the required workspace size and the computational load within the contraction kernel.

\subsection{Workspace Buffer Size Estimation}
\label{sec:workspace-buffer-size}

The workspace buffer size depends on two main factors: 
(1) the required space for scheduler metadata, and 
(2) the required space for storing partial split-k outputs.

\paragraph{Scheduler Metadata.} The maximum size of each metadata section is derived from the largest possible number of concurrent requests and the maximum accumulated request length. Users must provide these upper bounds during the scheduler's first planning stage.

\paragraph{Partial Outputs.} The size of partial outputs depends on both the problem dimensions (i.e., the number of heads and the head dimension) and the number of CTAs per kernel launch. In our load-balancing algorithm~\ref{sec:load-balanced-scheduling}, only requests deemed ``long'' -- those whose KV length exceeds the total KV length divided by the number of CTAs -- are split. According to the Writethrough Optimizations in Section~\ref{sec:writethrough-optimizations}, only these split requests produce outputs in the workspace buffer. Because the number of splits cannot exceed the total number of CTAs, and each split yields at most two tiles that must be merged, there are at most \(2 \times \#\text{CTA}\) partial outputs. Each tile produces a partial output of size \(T_q \cdot H_{qo} \cdot (D + 1) \), where \(T_q\) is the query tile size, \(H_{qo}\) is the number of heads, and \(D+1\) is the head dimension and LSE dimension. Therefore, the upper bound for the total partial output size is:
\[
  2\,\#\text{CTA} \times T_q \times H_{qo} \times (D + 1).
\]
By default, the total number of CTAs is set to $k \times \#\text{SM}$, where $\#\text{SM}$ denotes the number of streaming multiprocessors on the GPU and $k$ is chosen to maximize CTA-level occupancy.
For tensor-core based microkernels with high register usage, $k$ typically does not exceed 2 on Ampere, and it is often 1 on Hopper (one CTA per SM, also referred to as a persistent kernel).

\section{Overlap of Attention with Other Operations}
\label{sec:overlap-attention}

Nanoflow~\cite{zhu2024nanoflow} overlaps GEMM, attention, and inter-device communication in separate CUDA streams, assigning a fixed number of SMs to each operation. In \sys, this SM number can be provided by the user through the plan functions, and the \sys load-balancing scheduler will allocate tiles accordingly.

\section{FP8--FP16 Mixed-Precision Attention}
\label{sec:mixed-precision}

Recent LLMs frequently adopt \texttt{fp8} KV-Cache to reduce memory bandwidth and storage costs~\cite{micikevicius2022fp8}. In \sys, we implement \emph{mixed-precision} attention kernels wherein the query and output remain in \texttt{fp16}, while the KV-Cache is stored in \texttt{fp8}. We leverage the fast numerical array converter and fragment shuffler proposed by \citet{gupta2024mixed} to accelerate dequantization and handle bitwidth mismatches efficiently. This design allows for reduced memory footprints and higher bandwidth utilization without significantly compromising numerical accuracy.

\section{Additional Evaluation}
\label{sec:additional-evaluation}

In this section, we present additional evaluation results to further validate the performance, scalability, and robustness of \sys across diverse experimental conditions.

\subsection{Comparison with FlexAttention}

We compare \sys and FlexAttention~\cite{he2024flexattention} on different attention variants using the AttentionGym~\cite{AttentionGym} benchmark on NVIDIA H100 80GB SXM.
We evaluated with batch size $16$, number of heads $16$ and head dim $128$, the CUDA version and the Triton version were fixed to 12.4 and 3.2, respectivelyrespectively.
Tables \ref{table:flexattention-causal-attn} to \ref{table:flexattention-sliding-window} show the performance of \sys and FlexAttention in TFLOPS/s, where higher numbers mean better performance.
Across all four scenarios and a range of sequence lengths, \sys consistently outperforms FlexAttention, with especially large gains at longer sequence lengths.
The better performance is mainly due to the usage of Hopper microarchitecture’s advanced features (such as warp specialization and TMA), and CUTLASS’s fine-grained resource control (at register-level rather than tile-level) over Triton. Note that these gaps will be alleviated once Triton fully supports these features.

\begin{table}[h]
\centering
\caption{Causal Attention}
\label{table:flexattention-causal-attn}
\begin{tabular}{lrr}
\toprule
Seq Length & FlexAttention & \sys \\
\midrule
512   & 209.11 & \textbf{250.454} \\
1024  & 294.53 & \textbf{406.554} \\
2048  & 376.90 & \textbf{487.236} \\
4096  & 421.00 & \textbf{548.388} \\
8192  & 441.26 & \textbf{587.903} \\
16384 & 453.57 & \textbf{612.259} \\
\bottomrule
\end{tabular}
\end{table}

\begin{table}[ht]
\centering
\vspace{-10pt}
\caption{Attention with Logits SoftCap}
\label{table:flexattention-logits-softcap}
\begin{tabular}{lrr}
\toprule
Seq Length & FlexAttention & \sys \\
\midrule
512   & 241.51 & \textbf{336.487} \\
1024  & 327.50 & \textbf{409.534} \\
2048  & 379.57 & \textbf{468.769} \\
4096  & 403.39 & \textbf{489.667} \\
8192  & 407.82 & \textbf{515.573} \\
16384 & 409.89 & \textbf{520.935} \\
\bottomrule
\end{tabular}
\end{table}

\begin{table}[ht]
\centering
\caption{ALiBi Bias~\cite{press2022alibi}}
\label{table:flexattention-alibi}
\begin{tabular}{lrr}
\toprule
Seq Length & FlexAttention & \sys \\
\midrule
512   & 253.22 & \textbf{403.899} \\
1024  & 344.70 & \textbf{500.220} \\
2048  & 406.14 & \textbf{535.498} \\
4096  & 426.13 & \textbf{561.324} \\
8192  & 436.35 & \textbf{573.493} \\
16384 & 434.86 & \textbf{578.005} \\
\bottomrule
\end{tabular}
\end{table}

\begin{table}[ht]
\centering
\vspace{-3pt}
\caption{Sliding Window (window size = $1024$)}
\label{table:flexattention-sliding-window}
\begin{tabular}{lrr}
\toprule
Seq Length & FlexAttention & \sys \\
\midrule
512   & 206.51 & \textbf{236.363} \\
1024  & 292.25 & \textbf{374.108} \\
2048  & 350.91 & \textbf{381.464} \\
4096  & 368.45 & \textbf{384.998} \\
8192  & 373.25 & \textbf{384.514} \\
16384 & 367.91 & \textbf{380.506} \\
\bottomrule
\end{tabular}
\end{table}

\subsection{Evaluation of Shared-Prefix Attention Kernels}

We measure shared-prefix attention kernels with suffix length $128$.
Table~\ref{table:eval-shared-prefix-attn} shows the kernel latency under different shared prefix lengths, scenarios and batch sizes, where numbers are in microseconds (us), and ``composable'' means composable format while ``single'' means single format.
The composable format benefits long prefixes (e.g., 32k) and large batch sizes (e.g., $64$).
However, these speedups do not always yield proportional end-to-end gains because real-world shared prefix sizes tend to be smaller.

\begin{table}[hbt]
\small
\centering
\caption{Latency of Shared-Prefix Attention Kernels}
\label{table:eval-shared-prefix-attn}
\begin{tabular}{lr@{}r@{}r@{}r@{}}
\toprule
\makecell{Shared \\ Prefix \\ Length} & \makecell{Composable \\ (BS=16)} & \makecell{Single \\ (BS=16)} & \makecell{Composable \\ (BS=64)} & \makecell{Single \\ (BS=64)} \\
\midrule
1024   & \textbf{45.17}  & 46.52  & 87.86   & 130.49 \\
8192   & \textbf{88.67}  & 226.57 & 125.76  & 931.75 \\
32768  & \textbf{217.42} & 945.67 & 254.54  & 4090   \\
\bottomrule
\end{tabular}
\end{table}

\subsection{Ablation Study on Variable Sequence Length and load-balancing scheduler}

We conduct ablations on the effect of load-balancing scheduler~(Section~\ref{sec:load-balanced-scheduling}).
Table~\ref{table:eval-ablation-itl} and~\ref{table:eval-ablation-ttft} show the results for Llama 3.1-8B-Instruct running on an NVIDIA H100 SXM5 GPU with SGLang~\cite{zheng2023sglang} + \sys (with and without load-balancing scheduler).
We evaluate the inter-token latency (ITL, ms) and time-to-first-token (TTFT, ms) with three datasets: ShareGPT, variable sequence length with input lengths sampled from $U(512, 2048)$ and output fixed at $256$, and variable sequence length with input lengths sampled from $U(4096, 16384)$ and output fixed at $256$. ``RR'' in the tables means request rate.

\begin{table}[hbt]
\small
\centering
\vspace{-5pt}
\caption{Load-balancing Scheduler Ablation Study (ITL)}
\label{table:eval-ablation-itl}
\begin{tabular}{@{}lrrr@{}}
\toprule
Scenario & \makecell{w/ \\ Load- \\ Balancing} & \makecell{w/o \\ Load- \\ Balancing} & Triton \\
\midrule
ShareGPT (RR=16) & \textbf{8.96} & 9.16 & 9.36 \\
$U(512, 2048)$ (RR=8)        & \textbf{8.21} & 8.42 & 8.49 \\
$U(4096, 16384)$ (RR=1)        & \textbf{8.63} & 13.89 & 11.08 \\
\bottomrule
\end{tabular}
\end{table}

\begin{table}[hbt]
\small
\centering
\caption{Load-balancing Scheduler Ablation Study (TTFT)}
\label{table:eval-ablation-ttft}
\begin{tabular}{@{}lrrr@{}}
\toprule
Scenario & \makecell{w/ \\ Load- \\ Balancing} & \makecell{w/o \\ Load- \\ Balancing} & Triton \\
\midrule
ShareGPT (RR=16) & \textbf{39.05} & 39.42 & 52.92 \\
 $U(512, 2048)$ (RR=8)        & \textbf{66.78} & 67.38 & 68.48 \\
 $U(4096, 16384)$ (RR=1)        & \textbf{411.02} & 421.60 & 566.30 \\
\bottomrule
\end{tabular}
\end{table}

\subsection{vLLM Integration Evaluation}

We compare the vLLM with \sys backend and its default backend with a fixed request rate of $16$, reporting throughput (tokens/s), inter-token latency (ITL, ms), and time-to-first-token (TTFT, ms) in Table~\ref{table:eval-vllm-integration}.
\sys reduces ITL by aroudn 13\% using fp8 KV-cache, but heavy Python overhead in vLLM integration (e.g. array operations) at host side causes minor regressions with bf16. Our future optimizations will address these in C++ and move the scheduler to device.

\begin{table}[hbt]
\centering
\caption{vLLM Integration Evaluation}
\label{table:eval-vllm-integration}
\begin{tabular}{@{}l@{}rrr@{}}
\toprule
Backend         & Throughput & \makecell{Median \\ ITL} & \makecell{Median \\ TTFT} \\
\midrule
Default (bf16)  & 6062.89    & 10.42      & 35.85      \\
\sys (bf16)    & 6065.41    & 10.63      & 36.60      \\
Default (e4m3)  & 6015.86    & 12.56      & 39.74      \\
\sys (e4m3)    & 6020.32    & 10.92      & 37.93      \\
\bottomrule
\end{tabular}
\end{table}

\subsection{Fine-Grained Block-Sparsity Evaluation}

\sys supports fine-grained block-sparse matrices, which is useful in many KV-Cache pruning algorithms.
We measure the kernel performance on Quest~\cite{tang2024quest}, a state-of-the-art long-context modeling algorithm, which uses fine-grained sparsity in KV-Cache.
We compared the batch decoding attention kernel in Quest using \sys and compared its performance to PyTorch SDPA and FlexAttention on an NVIDIA H100 SXM5 GPU with the configuration (block size 16, num\_qo\_heads $32$, num\_kv\_heads $32$, head\_dim $128$).
All latency values reported are in microseconds (us).

As shown in Table~\ref{table:eval-sparsity-flashinfer} to~\ref{table:eval-sparsity-flexattention}, \sys demonstrates a considerable performance advantage, achieving up to a 20x speedup for long sequence lengths.
Currently FlexAttention relies on large block size templates, while \sys employs a sparse-row gathering strategy to leverage dense tensor cores for small block sizes. This design choice supports fine-grained KV-cache pruning.

\begin{table}[hbt]
\small
\centering
\caption{\sys Fine-Grained Sparsity  Latency (us)}
\label{table:eval-sparsity-flashinfer}
\begin{tabular}{@{}lrrrr@{}}
\toprule
\multicolumn{1}{c}{} & \multicolumn{4}{c}{page\_budget} \\
\cmidrule(l){2-5}
seq\_len & 64 & 128 & 256 & 512 \\
\midrule
4096  & 20.299 & 30.361 & 44.383 & 44.430 \\
8192  & 22.273 & 28.603 & 44.928 & 68.194 \\
16384 & 20.485 & 28.678 & 44.677 & 68.700 \\
32768 & 22.371 & 28.700 & 44.988 & 68.478 \\
\bottomrule
\end{tabular}
\end{table}

\begin{table}[hbt]
\small
\centering
\caption{PyTorch SDPA Fine-Grained Sparsity Latency (us)}
\label{table:eval-sparsity-torch}
\begin{tabular}{@{}lrrrr@{}}
\toprule
\multicolumn{1}{c}{} & \multicolumn{4}{c}{page\_budget} \\
\cmidrule(l){2-5}
seq\_len & 64 & 128 & 256 & 512 \\
\midrule
4096  & 287.684 & 288.904 & 287.715 & 287.807 \\
8192  & 474.631 & 474.508 & 474.683 & 473.070 \\
16384 & 857.319 & 857.570 & 857.094 & 857.728 \\
32768 & 1711.955 & 1711.621 & 1713.093 & 1711.709 \\
\bottomrule
\end{tabular}
\end{table}

\begin{table}[hbt]
\small
\centering
\caption{FlexAttention Fine-Grained Sparsity Latency (us)}
\label{table:eval-sparsity-flexattention}
\begin{tabular}{@{}lrrrr@{}}
\toprule
\multicolumn{1}{c}{} & \multicolumn{4}{c}{page\_budget} \\
\cmidrule(l){2-5}
seq\_len & 64 & 128 & 256 & 512 \\
\midrule
4096  & 1100.349 & 1097.356 & 1073.753 & 1071.797 \\
8192  & 1092.695 & 1099.100 & 1078.081 & 1074.886 \\
16384 & 1109.817 & 1101.535 & 1077.639 & 1076.859 \\
32768 & 1169.109 & 1187.395 & 1176.332 & 1174.502 \\
\bottomrule
\end{tabular}
\end{table}

%%%%%%%%%%%%%%%%%%%%%%%%%%%%%%%%%%%%%%%%%%%%%%%%%%%%%%%%%%%%%%%%%%%%%%%%%%%%%%%
%%%%%%%%%%%%%%%%%%%%%%%%%%%%%%%%%%%%%%%%%%%%%%%%%%%%%%%%%%%%%%%%%%%%%%%%%%%%%%%

\end{document}